# Nanoscale austenite reversion through partitioning, segregation, and kinetic freezing: Example of a ductile 2 GPa Fe-Cr-C steel

L. Yuan[1], D. Ponge[1], J. Wittig[1,2], P. Choi[1], J. A. Jiménez[3], D. Raabe[1]

[1] Max-Planck-Institut für Eisenforschung, Max-Planck-Str. 1, 40237 Düsseldorf, Germany
[2] Vanderbilt University, Nashville, TN 37235-1683, USA
[3] CENIM-CSIC, Avda. Gregorio del Amo 8, 28040-Madrid, Spain

## Abstract

Austenite reversion during tempering of a Fe-13.6Cr-0.44C (wt.%) martensite results in an ultra-high strength ferritic stainless steel with excellent ductility. The austenite reversion mechanism is coupled to the kinetic freezing of carbon during low-temperature partitioning at the interfaces between martensite and retained austenite and to carbon segregation at martensite-martensite grain boundaries. An advantage of austenite reversion is its scalability, i.e., changing tempering time and temperature tailors the desired strength-ductility profiles (e.g. tempering at 400°C for 1 min. produces a 2 GPa ultimate tensile strength (UTS) and 14% elongation while 30 min. at 400°C results in a UTS of ~ 1.75 GPa with an elongation of 23%). The austenite reversion process, carbide precipitation, and carbon segregation have been characterized by XRD, EBSD, TEM, and atom probe tomography (APT) in order to develop the structure-property relationships that control the material's strength and ductility.

## 1. Introduction

A high demand exists for lean, ductile, and high strength Fe-Cr stainless steels in the fields of energy conversion, mobility, and industrial infrastructure. As conventional martensitic stainless steels (MSS) typically exhibit brittle behavior, supermartensitic Fe-Cr stainless steels (SMSS) with enhanced ductility have been designed in the past years by reducing carbon (<0.03 wt.%) and adding nickel (4%-6.5 wt.%) and molybdenum (2.5 wt.%) [1-4]. The heat-treated microstructures of these materials are characterized by tempered martensite and retained austenite[1-4].

In this work we present an alternative approach of designing MSS steels with both, high strength and ductility. Our method is based on nanoscale austenite reversion and martensite relaxation via a modest heat treatment at 300-500°C for several minutes. We make the surprising observation that this method leads to very high strength (up to 2 GPa) of a Fe-13.6Cr-0.44C (wt.%) steel without loss in ductility (X44Cr13, 1.4034, AISI 420).

Quenching followed by tempering is known to improve the strength and toughness of martensitic steels[5-7]. Specifically, quench and partitioning (Q&P) treatments are efficient for producing steels with retained austenite and improved ductility[8]. The heat treatment sequence for Q&P steel involves quenching to a temperature between the martensite-start ($M_s$) and martensite-finish ($M_f$) temperatures, followed by a partitioning treatment either at, or above, the initial quench temperature. Partitioning is typically designed in a way to enrich and stabilize the retained austenite with carbon from the supersaturated martensite [9]. In conventional Q&P processes, the quench temperature is hence chosen in a way that some retained austenite prevails and subsequent tempering leads to carbon partitioning between martensite and austenite. Typically, no new austenite is formed during partitioning.

In our study we modify this approach with the aim to increase the amount of austenite during low-temperature partitioning. We start with austenitization and water quenching to room temperature. This provides a martensitic-austenitic starting microstructure. During a subsequent heat treatment in the range 300°C-500°C, austenite reversion[10-15] takes place on the basis of partial partitioning according to local equilibrium, segregation, and kinetic freezing of carbon inside the newly formed austenite.

It is important to point out that the phenomena occurring during austenite reversion are in the present case different from conventional Q&P approaches: In Q&P processing, the carbon diffuses from martensite into the already present austenite during tempering where equilibration of the carbon distribution inside the austenite is generally assumed. In the current case of low-temperature partitioning, however, the carbon is enriched in front of the austenite boundary and accumulates there since it has a much higher diffusion rate in bcc than in fcc. The accumulated carbon at the martensite-austenite interface than provides a high local driving force for austenite reversion. Once captured by the growing austenite, the carbon is kinetically frozen owing to its small mobility in fcc. The phenomena occurring during austenite reversion in Fe-Cr-C stainless steels are complex due to the high content of carbon and substitutional alloying elements. In contrast to typical Q&P steels where carbide precipitation ($M_3C$) is suppressed by alloying with Si and/or Al[16], in the present alloy $M_3C$-type carbide precipitation occurs at 400°C. This means that a kinetic and thermodynamic competition exists for carbon between austenite reversion, enrichment of retained austenite, and carbide formation during tempering.

Therefore, the partitioning temperature must be chosen on a theoretically well founded basis for two reasons: First, low temperature annealing requires more local carbon enrichment to provide a driving force high enough for austenite reversion. We emphasize in this context that the local equilibrium matters for this process, i.e. a high carbon content is required at the martensite-austenite interface (not everywhere within the austenite). Equilibration of the carbon inside of the austenite is not necessarily required. Second, high temperature annealing may cause more carbide formation, consuming too much carbon, so that austenite reversion is suppressed due to an insufficient carbon chemical potential to promote it.

In order to elucidate the competing phenomena occurring during such low-temperature partitioning, namely, carbide formation vs. austenite reversion as well as the carbon redistribution inside the retained and reversed austenite fractions, atom probe tomography (APT) was used. This method allows us to measure the carbon content inside the austenite, which determines its stability, as well as inside the martensite and the carbides [17-28]. The APT method allows for three-dimensional elemental mapping with nearly atomic resolution and provides information about internal interfaces and local chemical gradients[28-32].

## 2. Experimental

The material used in this study was a martensitic stainless steel with the chemical composition Fe-13.6Cr-0.44C (wt.%; 1.4034, X44Cr13, AISI 420) which was provided by ThyssenKrupp Nirosta as a cold rolled sheet, table 1. The $Ae_3$ temperature, calculated by Thermo-Calc[33] using the TCFE5 data base[34], indicates that the incipient holding temperature for full austenitization should be above 800°C. The calculation further reveals that full dissolution of chromium carbides in austenite is achieved at about 1100°C. Hence, the annealing conditions were set to 1150°C for 5 minutes. Dilatometer tests were performed using a Bähr Dil805 A/D quenching and deformation device to identify the $M_s$ temperature during quenching. After water quenching, tempering at 300°C, 400°C, and 500°C, respectively, with different holding times was performed to study carbon redistribution, austenite reversion, and carbide formation (Fig. 1).

Mechanical properties were determined by tensile and Vickers hardness measurements (980N load, HV10). Tensile tests were carried out along the rolling direction of the samples at room temperature. Flat tensile specimens were machined with a cross section of 2.5mm x 8mm and a gauge length of 40mm. The tests were conducted on a Zwick/Roell Z100 tensile testing machine at a constant cross head speed of 1mm/min, corresponding to an initial strain rate of $4.2\times10^{-4}s^{-1}$.

The volume fraction of the austenite phase after heat treatments (carbide dissolution annealing and tempering at 400ºC for 1, 2, 10 and 30 minutes) was measured by x-ray diffractometry (XRD), electron back scattering diffraction (EBSD), and magnetic characterization (Feritscope MP30E-S).

EBSD samples were prepared by standard mechanical grinding and polishing procedures normal to the rolling direction. Subsequently, these samples were electropolished using Struers electrolyte A3 at room temperature using a voltage of 40V, a flow rate of 20/s and a polishing time of 20 s. EBSD was performed on a JEOL-6500F high-resolution field-emission scanning electron microscope operated at 15 kV [35].

X ray diffraction (XRD) measurements were carried out using $Co_{K\alpha}$ radiation. XRD data were collected over a 2θ range of 30-138º with a step width of 0.05º and a counting time of 10 s/step. The Rietveld method was used for the calculation of the structural parameters from the diffraction data of the polycrystalline bulk materials. We used version 4.0 of the Rietveld analysis program TOPAS (Bruker AXS). The analysis protocol included consideration of background, zero displacement, scale factors, peak breath, unit cell parameter, and texture parameters. The room temperature structures used in the refinement were martensite/ferrite and austenite.

Thin foils were prepared using standard twin-jet electropolishing from the as-quenched material and the tempered samples before and after deformation[36]. These samples were examined in a Philips CM 20 transmission electron microscope (TEM) at an acceleration voltage of 200 kV to characterize the carbide evolution and the formation of reverted austenite. Carbide characterization was also carried out by using a carbon extraction replica technique[37] and investigated by electron diffraction and energy dispersive spectroscopy (EDS) in the TEM.

Needle-shaped APT samples were prepared applying a combination of standard electropolishing and subsequent ion-milling with a focused-ion-beam (FIB) device. APT analyses were performed with a local electrode atom probe (LEAP™ 3000X HR) in voltage mode at a specimen temperature of about 60 K. The pulse to base voltage ratio and the pulse rate were 15% and 200 kHz, respectively. Data analysis was performed using the IVAS software (Cameca Instruments).

## 3. Results

### 3.1 Mechanical properties

The as-received cold rolled and recrystallized material has an ultimate tensile strength (UTS) of 640 MPa and a uniform elongation of 19%, Fig. 2a. After austenitization at 1150°C and water quenching, the material is brittle and fails before the yield stress is reached at a stress of 400 MPa (Fig. 2a). Thus, the true UTS for the as quenched state could be only estimated from the indentation hardness. The relationship between Vickers hardness (HV) and tensile strength was calculated considering a linear relationship of the form $\Delta HV = K \times \Delta UTS$. A constant K of 3.5 was determined by linear regression through data obtained from the hardness and UTS values obtained from samples after tempering at 400ºC for different times. Fig 2e suggests that the hardness for the as-quenched state corresponds to a tensile strength of more than 2300 MPa.

Fig. 2a also shows the stress-strain curves obtained from the tensile tests performed on samples tempered at 400ºC for different times. The most remarkable feature of these curves is the transition from a brittle behavior in the as quenched material to a ductile one after tempering. When the tempering time is increased, we observe an increase in uniform elongation and a decrease in UTS. After 30 minutes, the uniform elongation of the sample reaches a value of about 22% and a UTS above 1760 MPa. This value for the UTS can be also reached upon tempering at 500°C, but in this case, a gradual increase in total elongation upon increase in tempering time is not observed, Fig 2b. It can be observed that in this case the stress does not go through a maximum; that is, $\partial\sigma/\partial e$ (the partial derivative of the stress with respect to strain) does not go through zero. This would indicate that the sample fractures before the strain reaches the necking value. At 300°C, after 1 minute, the ductility improves slightly, i.e. longer tempering is required for obtaining better ductility at this temperature, as shown in Fig. 2c. When comparing the mechanical properties of samples tempered at different temperatures (Fig. 2d), the 400 °C treatment yields the optimum improvement in both UTS and total elongation (TE).

### 3.2 Phase fractions and kinetics: predictions and experiments

Thermo-Calc was used to calculate the phase equilibrium at the different partitioning temperatures. For evaluating kinetics during heating and cooling, we conducted dilatometer tests (Fig. 3). The heating and cooling rates were set to 10 K/s and -30 K/s, respectively. Above 876°C, the microstructure is fully austenitic. The $M_s$ temperatures were derived from the dilatometer tests (118°C after 1150°C annealing and 360°C after 950°C annealing). The Thermo-Calc calculations were used to predict the equilibrium carbon content of the austenite after annealing at different temperatures (Fig. 3c).

Fig. 4 shows the phase fraction of austenite versus tempering time for 400°C measured by feritscope (magnetic signal), EBSD, and XRD. For the as-quenched state the EBSD result provides a higher volume fraction (20%) than the magnetic (14.5%) and the XRD data (8%) which is attributed to the limited statistics of the EBSD method. During the first 2 minutes of tempering the amount of austenite increases rapidly, indicating austenite reversion. After 30 minutes, nearly 40 vol.% austenite is observed consistently for all three methods.

Fig. 5 shows in situ EBSD observations of the austenite during tempering. Fig. 5(a) maps the material in the as-quenched state containing only retained austenite. Fig. 5(b) shows the same area during the in-situ experiment containing both, retained plus reverted austenite after 5 minutes tempering at 400°C subsequent to the quenching treatment. The EBSD map reveals the fine dispersion of the newly formed reverted austenite after 5 minutes. We observe two kinds of austenite, namely, one with a coarse topology and another one with a fine and dispersed topology.

The microstructures of the samples tempered for 0, 1 and 2 minutes, respectively, are shown in Fig. 6. The as-quenched material (0 minute tempering) is brittle and failed already in the elastic regime during tensile testing. From the microstructure it can be seen (Fig. 6b, left: before tensile test; right: after tensile test, for each state) that only a small amount of austenite was transformed to martensite when the material failed, i.e. austenite bands can still be observed near the fracture interface. For samples after 400°C tempering, no premature failure takes place and the total elongation (TE) reaches 14% (engineering strain). The microstructure at the fracture zone shows nearly no remaining austenite. This observation indicates that deformation-driven austenite-to-martensite transformation takes place. Secondary cracks along the tensile direction are visible in the EBSD maps. It seems that these cracks follow the band-like former retained austenite regions, which transformed during straining into martensite.

3.3 TEM characterization

After solid solution and subsequent water quenching, we found no retained austenite in the TEM foils (Figs. 7a,b). This is in contrast to the results obtained from the EBSD maps which show retained austenite in the as-quenched state (Figs. 5,6). We attribute this discrepancy between TEM and EBSD results to the fact that the as-quenched metastable retained austenite - when thinned for TEM analysis - is no longer constrained by the surrounding martensite and hence transforms into martensite.

After 1 minute tempering at 400°C we observe a high tensile strength of 2 GPa, Fig. 2a. The corresponding microstructure was monitored by TEM, Fig. 7. Figs. 7c,d give an overview of the nanoscaled elongated carbides formed during tempering.

The carbides have an average length of 70nm and an average width of 5nm. After 30 minutes tempering the average particle spacing is about 80nm and the length 110 nm. The carbides after 1 minute tempering at 400°C were examined via carbon extraction replica. The diffraction patterns reveal that they have $M_3C$ structure. This means that the formation of $M_{23}C_6$ carbides is suppressed at such a low tempering temperature. Energy Dispersive Spectroscopy (EDS) analyses showed that the metal content in the carbide (M in $M_3C$) amounts to 74 at.% Fe and 26 at.% Cr, i.e. the Cr/Fe atomic ratio is 0.35. The measured chromium content in the $M_3C$ carbides significantly deviates from the nominal chromium concentration of 14.2 at.% Cr / 82.5 at.% Fe = 0.17.

Fig. 7e shows the formation of a thin austenite layer that is located at a former martensite-martensite grain boundary. Fig. 7f is a close-up view of a thin austenite zone that is surrounded by martensite. Electron diffraction analysis reveals that a Kurdjumov-Sachs orientation relationship exists between the martensite matrix and the thin austenite layer, Fig. 7g [38]. In line with the in-situ EBSD results in Fig. 5, where we observed reverted austenite formed between martensitic grains, the austenite film observed here in TEM might be either retained or reversed austenite. In order to determine more reliably which of the two kinds is observed local atomic scale chemical analysis is conducted by using APT as outlined below. The two types of austenite can then be distinguished in terms of their carbon content: Retained austenite has at first the nominal quenched-in C content (about 2 at.% in the present case) of the alloy while reverted austenite has a higher C content (up to 9 at.%) owing to local partitioning and kinetic freezing. However, we also have to account for the possibility that the retained austenite can have a higher C content as the lath martensite mechanism is slow enough to allow for some C diffusion out of the martensite into the retained austenite during quenching.

### 3.4 Atom probe tomography

The local chemical compositions and their changes during 400°C tempering of the martensite, austenite, carbides, and interface regions were studied by atom probe tomography. Phase identification is in all cases achieved via the characteristic carbon contents of the present phases. Fig. 8 shows the 3D atom maps after water quenching (Fig. 8a), water quenching plus tempering at 400°C for 1 min (Fig. 8b), and water quenching plus tempering at 400°C for 30 minutes (Fig. 8c). Carbon atoms are visualized as pink dots and carbon iso-concentration surfaces in green for a value of 2 at.%. This value corresponds to the nominal carbon concentration of the alloy of 0.44 wt.%. The different phases (martensite, austenite, carbide) are marked. They were identified in terms of their characteristic carbon content and the TEM and EBSD data presented above. For more quantitative analyses, one-dimensional compositional profiles of carbon across the martensite-martensite and martensite-austenite interfaces were plotted (along cylinders marked in yellow in the 3D atom maps).

#### 3.4.1 As-quenched condition

Fig. 8a reveals that in the probed volume carbon is enriched along the martensite-austenite interface. The interface region, shown as composition profile in Fig. 8a, reveals an average carbon concentration of about 1.90 at.% in the austenite with strong local variations and of about 0.98 at.% in the abutting supersaturated martensite. The carbon concentration in the austenite nearly matches the nominal carbon concentration of the alloy. Some carbon clusters occur in both phases. The carbon concentration in these clusters is about 3 at.%, i.e. they are not carbides. In a thin interface layer of only about 5 nm, the carbon content is very high and reaches a level of 4-6 at.%. In contrast to the variation in the carbon distribution, the chromium content is the same in the martensite, the interface, and the austenite, Fig. 8a.

3.4.2 400°C tempered condition after quenching

After 1 minute tempering at 400°C, a carbon enriched austenite layer (15-20 nm width) is observed between two abutting martensite regions (Fig. 8b). The thin austenite zone contains in average about 6.86 at.% carbon while the martensite matrix contains only about 0.82 at.% carbon. The identification of the phases in these diagrams follows their characteristic carbon content.

After 30 minutes tempering (Fig. 8c), different carbon enriched areas appear. They correspond to individual phases. The analyzed volume can be divided into 2 zones. The top region with low carbon content corresponds to martensite. The bottom zone with higher carbon content corresponds to austenite. Inside the martensitic region there are areas with very high carbon content (see arrow in Fig. 8c). The carbon content is 25.1 at.% in this particle indicating $M_3C$ cementite stoichiometry. In the martensitic matrix surrounding the precipitate the carbon content amounts to only 0.48 at.%. Carbon partitioning to the different phases can be quantified in terms of an enrichment factor ε=(at.% C tempered)/(at.% C as quenched) to compare the compositions in the phases before and after tempering. The observed values of ε for each state are listed in Table 2. The carbon content in the martensite decreases continuously during tempering, which can be ascribed to carbon partitioning from the super-saturated martensite to the austenite and to carbide formation [39-50].

## 4. Discussion

### 4.1 Mechanisms of partitioning and austenite reversion

The microstructure observed by EBSD and TEM allows us to monitor the austenite development at the mesoscopic scale: during the initial high temperature solution annealing in the austenitic regime (1150°C for 5 minutes), all carbides were dissolved (Fig. 3c). The high content of solute carbon that is present in the austenite after carbide dissolution decreases the $M_s$ and the $M_f$ temperatures of the austenite below room temperature. Hence, 8-20 vol.% retained austenite exists after quenching the solution annealed material to room temperature, Fig. 4. The differences in retained austenite are due to the individual precision of the different characterization methods. EBSD provides a direct method and hence is assumed to give a realistic value within its statistical limits.

After tempering at 400°C for 30 minutes the area fraction of austenite increases to about 40%. This change documents that strong austenite reversion takes place even at this low temperature. The local variations in the austenite dispersion after short tempering were larger compared to longer tempering times. We attribute this heterogeneity in the re-austenitization kinetics and topology to the mean diffusion range of the carbon and to the distribution of the carbon sources. Using the data of Speer et al. [40] for the diffusion coefficients in ferrite $D_{\alpha}= 2\times10^{-12}$ $m^2/s$ and in austenite $D_{\gamma}= 5\times10^{-17}$ $m^2/s$ we obtain a mean free path for carbon of $1.5\times10^{-4}$ m in ferrite and $7.4\times10^{-7}$ m in austenite at 400°C and 30 minutes.

This means that austenite reversion starts at decorated defects (e.g. internal interfaces) where the local carbon concentration is high enough and the nucleation energy low enough to promote the formation of this phase. Fig. 7e confirms this assumption. The TEM analysis also suggests that austenite reversion proceeds via a Kurdjumov-Sachs orientation relationship. Shtansky et al.[38] found the same crystallographic relationship during reverse transformation in an Fe–17Cr–0.5C tempered martensite (wt.%).

An important aspect of the pronounced austenite reversion in the current case is that the competing formation of $M_{23}C_6$ carbides is suppressed at 400°C. This means that more carbon is available to stabilize and promote austenite formation [38,45-47].

We used Thermo-Calc predictions[33,34] to estimate the driving force for austenite reversion for the current alloy and the employed tempering conditions, Fig. 9b. The results reveal that if the carbon concentration in the bulk martensite (α') exceeds 1.21 wt.% (5.45 at.%), austenite (γ) will form at 400°C, provided that the nucleation barrier is overcome. This result confirms our suggestion made above, namely, that no bulk austenitization can occur at this temperature since the average carbon content of the matrix is too low. Instead we assume that only certain lattice defects (interfaces) that experience very high elastic distortions and carbon segregation can provide the nucleation conditions and a sufficiently high carbon concentration for local austenite formation, Fig. 7e,f,g, Fig. 8. This leads to an increase in the overall austenite fraction. Fig. 6a shows that a 2 minutes heat treatment at 400°C leads to an increase in the austenite content from 18.9 vol.% to 29.7 vol.%.

Thermo-Calc predictions show that in the current alloy carbon provides the required driving force for this low-temperature austenite reversion. Substitutional atoms, particularly Cr, do not participate in reversion in the current alloy owing to their limited mobility at 400°C, i.e. the driving force for transformation is here provided exclusively by the high carbon enrichment rather than by substitutional depletion of the austenite, Fig. 8a [39].

Based on these thermodynamic boundary conditions the APT results allow us now to monitor and evaluate the kinetics of carbon at different stages of tempering in more detail (Fig. 8). Fig 8a shows the as quenched state: During solution annealing where the material is completely austenitic the elements distribute homogeneously within that phase.

At the onset of water quenching, the majority of the austenite starts to transform into martensite without at first changing its chemical composition. However, as the solubility of carbon in the quenched-in martensite is very small, carbon starts to leave the martensite during and after the $\gamma \rightarrow \alpha'$ transformation and enriches at the $\gamma$-$\alpha'$ interfaces, Fig. 8a [45.-50]. This process can happen extremely fast: Speer et al. [40,41] showed that carbon partitioning between martensite and austenite in a 0.19C-1.59Mn-1.63Si (wt.%) steel required at 400°C less than 0.1s owing to the relatively high diffusion rate of carbon in martensite. In contrast, the further distribution of the newly acquired carbon within the austenite is nearly three orders of magnitude more slowly [40-50]. This means that in this case the escape rate of carbon from the newly forming lath martensite is much higher than the carbon equilibration within the austenite [48-50].

In the present quenching process, carbon segregation takes place even faster than in the study quoted above[40,41]. In the current case the carbon has already started to partition and segregate at the martensite-austenite interfaces during the early stages of water quenching immediately after the first martensite has formed[45-50]. The fast kinetics is due to the high mobility of carbon in martensite, table 3. Such pronounced carbon segregation at the martensite-austenite interfaces is clearly observed in the as quenched state (Fig. 8a). In the interface area the carbon content reaches up to 4-6 at.% within a narrow layer of about 5 nm. This value is clearly above the nominal composition that purely retained austenite would have[45-50]. Owing to the high escape rate of carbon from the martensite this zone is interpreted as a portion of initially retained austenite which has been enriched in carbon during quenching [45-50].

As explained above this high level of carbon segregation in the present case is a consequence of two effects, namely, first, the rapid carbon escape from the newly formed martensite and second, the low mobility of carbon within the retained austenite. According to table 3 at 400°C in 1 minute carbon can diffuse 27.000 nm in the martensite and only 130 nm in the austenite[40]. Other sources suggest a 10-20% smaller mean free path of carbon in the martensite [43,44].

Hence, the carbon segregation observed after quenching (Fig. 8a) is due to a partitioning step and a kinetic freezing step (limited mobility of carbon once it arrives in the austenite). From comparing this experimentally observed frozen-in value of 4-6 at.% carbon at the martensite-austenite interface (Fig. 8a) with the value that is predicted by Thermo-Calc as a driving force required for austenite reversion at 400°C (5.45 at.%) we conclude that austenite reversion will occur under the current conditions at this interface upon heat treatment.

After 1 minute tempering at 400°C, a carbon enriched austenite layer is observed between two martensite regions (Fig. 8b). In principle this thin austenite layer could originate either from a very thin layer of retained austenite that was enriched with carbon due to partitioning from the abutting lath martensite or from austenite reversion without any preceding retained austenite. If the carbon-rich zone would be retained austenite the carbon profile in the austenite region would assume a 'V' type distribution. This type of concentration profile would be characterized by a high content at the two martensite-austenite interfaces and a low content in the center of the austenite layer (hence 'V'). Also, retained austenite would have the nominal composition, i.e. in the center of the retained austenite zone the carbon content should be 2 at.% or slightly above as in Fig. 8a. This type of carbon distribution is not observed though. Instead, Fig. 8b shows that the carbon profile assumes a '/\' shape within the austenite layer with a maximum carbon concentration above 8 at.%. It is hence plausible to assume that this profile is due to carbon segregation on a former martensite-martensite grain boundary according to the Gibbs adsorption isotherm. This means that during water quenching the carbon that is segregated at the martensite grain boundary has come from both sides. Therefore, the maximum carbon concentration revealed in Fig. 8b, which highly exceeds the equilibrium concentration that it would have had in the austenite, is in the center of the enrichment layer rather than at its rims. This means that during tempering austenite reversion starts in the center of this carbon enriched area, i.e. at the former martensite-martensite grain boundary. The resulting average carbon concentration in this reverted austenite grain is very high, namely, 6.86 at.% (with a maximum above 8 at.%). The fact that the pronounced '/\' shape of the carbon is preserved (frozen in) inside the austenite is due to the low mobility of carbon in austenite, table 3. These observations suggest that the carbon-rich zone in Fig. 8b is a newly formed austenite layer. If the carbon enriched area had been located between a martensite and an austenite grain, such as at the positions observed in Figs. 8a and c, the carbon atoms would have arrived only from one side, namely, from the martensite side (Fig. 9). The thickness of the newly formed reverted austenite layer in Fig. 8b is about 15 nm. With increasing tempering time, more reverted austenite is formed (Fig. 4,9).

In summary, the behavior of carbon in the current alloy can be described as follows: during quenching, carbon segregates to martensite-martensite grain boundaries (equilibrium segregation) or to martensite-retained austenite interfaces (partitioning plus kinetic freezing). In the first case (equilibrium segregation between two lath martensite zones) during tempering, these carbon enriched areas in the martensite revert to austenite when the driving force is high enough owing to the favorable nucleation barrier at the interfaces. In the second case (partitioning at retained austenite) the carbon enrichment leads to austenite growth according to local equilibrium. If the so reverted austenite is located at or in the vicinity of the austenite-martensite phase boundary, carbon can diffuse from the reverted austenite further into the retained austenite provided that the tempering time is long enough. This carbon enrichment stabilizes the retained austenite. Also, this effect makes it generally difficult to distinguish reverted austenite from retained austenite (Fig. 9). After 1 minute tempering, the reverted austenite has a high carbon content of 6.86 at.% (enrichment factor $\varepsilon$=3.61). With increasing tempering time, the diffusion of carbon from reverted austenite into retained austenite leads to an increase in the carbon concentration of the retained austenite. After 30 minutes tempering, the carbon concentration in the retained austenite increases to an average value of 2.42 at.% (Fig. 8c). If the diffusion distance to the nearest phase boundary is too far, e.g. inside the bulk martensite, the high concentration of carbon leads to the formation of carbides inside the martensite.

After 30 minutes tempering time, the carbon content in the carbides is 25.1 at.% as measured by APT. This value agrees with the stoichiometric content of carbon in $M_3C$ (25 at.%). Due to the carbon partitioning to austenite, austenite reversion, and the competing formation of carbides, the carbon content of the martensite continuously decreases during tempering. The amount of carbon in each phase before and after tempering is listed in Table 2 for the different stages. The other elements, for example chromium, have nearly the same content in both, austenite and martensite. This means that during 400°C tempering, medium range diffusion of carbon can be observed, but the substitutional elements only experience short distance diffusion. For all tempering conditions analyzed above we observe that not the nominal (global) but the local chemical potential of carbon directly at the martensite-austenite and martensite-martensite interfaces and the smaller nucleation energy at the interfaces determine the kinetics of austenite reversion. Similar trends were observed in maraging steels during aging [28,51,52].

In a thought experiment, assuming infinite mobility of the carbon when entering from martensite into austenite, the reversion would proceed more slowly owing to the smaller chemical driving force at the interface when carbon is distributed more homogeneously inside the austenite. In the current situations, however, carbon becomes trapped and highly enriched at the martensite-austenite interface owing to the partitioning and its low mobility within the austenite. This provides a much higher local driving force for austenite reversion. We refer to this mechanism as low temperature partitioning and kinetic freezing effect because the carbon is fast inside the martensite when leaving it but slow (and, hence, frozen) when entering the austenite. A similar effect was recently observed in Fe-Mn steels [28].

4.2 Transformation induced plasticity (TRIP) effect

During tensile testing, the volume fraction of austenite decreases not only in the quenched samples but also in the tempered samples (Fig. 6). When the brittle as-quenched sample failed at an early stage of loading (Fig. 2a, green curve), the amount of retained austenite had decreased from 18.9 to 10.8 vol.%. At failure most of the quenched-in martensite was still in the elastic regime. This means that stress-induced austenite-to-martensite transformation prevailed since the material took nearly no plastic strain until fracture. After 400°C austenite-reversion tempering, the ductility of the material improves drastically (Fig. 2). The EBSD results reveal that nearly all of the austenite transformed into martensite during tensile testing especially in the near-fracture zones (Fig. 6a,b). This observation suggests that strain-induced austenite-to-martensite transformation (rather than stress-induced transformation) prevails in the tempered samples containing reverted nano-sized austenite.

The difference in the displacive deformation behavior between the as-quenched and tempered samples is due to the fact that directly after water quenching, the retained austenite is unstable due to its relatively low carbon content. In the as-quenched state (i.e. without tempering) the carbon content of the retained austenite is equal to the nominal composition after solution treatment. A relatively weak load is, hence, required to transform this retained and rather unstable austenite into martensite at the onset of the tensile test. Transforming a large amount of austenite at the same time, namely, at the beginning of deformation, promotes crack formation and premature failure. In contrast to this as-quenched and rather unstable austenite, subsequent tempering enriches the retained austenite with carbon due to partitioning. The higher carbon content stabilizes the retained and the reverted austenite so that austenite portions with different carbon content undergo the TRIP effect at different stages of deformation. These differences in carbon content of different austenite portions in the same sample is due to the fact that only the local chemical potential of the carbon at the hetero-interfaces determines the partitioning and reversing rates and, hence, also the exact carbon content of the abutting austenite. This means that retained and reverted austenite zones that are carbon-enriched through partitioning have a kinetically determined composition which is subject to local variations in the chemical potential (of carbon). This context explains the more continuous displacive transformation sequence in the tempered material and hence the observed ductility improvement.

Another aspect of the TRIP effect in this material is that austenite reversion, obtained from tempering, does not only stabilize the austenite via a higher carbon content but also increases its overall volume fraction at least after sufficient tempering time. Fig. 6a reveals that the austenite fraction increases from 18.9 vol.% after quenching to 29.7 vol.% after 2 minutes at 400°C. Interestingly, after 1 minute at 400°C the austenite fraction did not change much. This means that for the short-annealing case (1 minute) the austenite stability and its more sequential transformation as outlined above are more important for the ductilization than its mere volume fraction.

### 4.3 Precipitation development

The TEM and APT observations confirm that the carbides formed during tempering have are of $M_3C$ type (instead of $M_{23}C_6$). The formation of $M_3C$ is associated with a smaller loss of chromium from the matrix (into carbides) compared to $M_{23}C_6$-type carbides which can have a high chromium content. Some authors found a sequence of carbide formation in Fe-Cr-C systems during tempering according to $MC \rightarrow M_3C \rightarrow M_7C_3 + M_{23}C_6 + M_6C$[53]. In our study $M_3C$ carbides prevailed up to 30 minutes annealing time. Samples taken from the as-quenched state show the highest hardness due to carbon in solid solution. The hardness decrease observed during tempering is related to carbide formation because carbon has a higher strengthening effect in solid solution than in the form of carbides. However, the small carbides (Fig. 8c) also contribute to the strength as observed with TEM (see Fig. 7b). The strain hardening rate decreases with increasing tempering time. This might be due to the coarsening of the carbides and due to the increase in the average carbide spacing (from ~40 nm after 1 minute to ~80 nm after 30 minutes tempering at 400°C). Further we observe that the yield stress increases during tempering. This might be due to the change in the internal stress state of the martensite matrix. After water quenching, high elastic stresses prevail in the martensite. These lead to early microplastic yielding of the material prior to percolative bulk plastic yielding. During tempering, the internal stress state of the martensite is relaxed due to the escape of carbon. This leads to a delay in the yielding of the tempered samples.

### 4.4 Relationship between nanostructure and stress-strain behavior

In the preceding sections we presented experimental evidence of grain boundary segregation, hetero-interface partitioning, kinetic freezing, carbide precipitation, retained austenite formation and stabilization, austenite reversion, and the TRIP effect.

In this part we discuss the joint influence of these phenomena on the excellent strength-ductility profile of this steel (Fig. 2a,d).

We differentiate between mechanisms that provide higher strength and those promoting ductility: The most relevant phase responsible for the high strength of the steel after heat treatment is the relaxed martensite. The quenched-in martensite with an extrapolated tensile strength of more than 2300 MPa (approximated from hardness data) is very brittle. Already a very modest heat treatment of 1 minute at 400°C though (Fig. 2a) provides sufficient carbon mobility. This leads to carbon re-distribution (carbide formation, grain boundary segregation, dislocation decoration, martensite-austenite interface segregation, austenite solution) and thus to a reduction in the internal stresses of the martensite. The reduced carbon content renders the martensite into a phase that can be plastically deformed without immediate fracture. The second contribution to the increase in strength are the nanoscaled carbides which provide Orowan strengthening, Fig. 7 (TEM), Fig. 8c (APT). Their average spacing increases from about 50 nm (1 minute at 400°C) to about 80nm (30 minute at 400°C), Fig. 8b. These two effects, viz., conventional martensite strength (via high dislocation density, high internal interface density, internal stresses, solid solution strengthening) and Orowan strengthening explain the high strength of the material, but they do not explain its high ductility.

In this context the TRIP effect, i.e. the displacive transformation of retained and reverted austenite, becomes relevant: Fig. 6a reveals a drop in the austenite content from 29.7 vol.% to about 2.7 vol.% during deformation for the sample heat treated at 400°C for 2 minutes. Fig. 10 shows the true stress-true strain curves and their corresponding derivatives (strain hardening) after 400°C heat treatment at different times. The data reveal that the tempering, which increases the austenite content via reversion, leads indeed to higher strain hardening reserves at the later stages of deformation due to the TRIP effect, Fig. 6b. Since longer heat treatments lead to higher volume fractions of reverted austenite the TRIP-related strain hardening assumes a higher level for these samples (Fig. 10).

Another important effect that might promote ductility in this context is the wide distribution of the austenite dispersion and stability (carbon content) which are both characteristic for this material. As revealed in Fig. 8 we can differentiate 3 types of austenite, Fig. 9a: The first type is the as-quenched retained austenite with the nominal carbon content and relatively low stability. The second one is retained austenite which assumes an increased carbon content due to partitioning during quenching and particularly during heat treatment and has thus higher stability against displacive transformation. The third type of austenite is the reverted one. These three types of austenite have different carbon concentration, volume fraction, and size. Both, carbon content and size affect austenite stability. This means that the displacive transformation during tensile testing and the associated accommodation plasticity occur more gradually upon loading compared to a TRIP effect that affects a more homogeneous austenite. We refer to this mechanism as a heterogeneous TRIP effect.

Another important aspect is the composite-like architecture of the reverted austenite, which is located at the martensite-martensite and at the martensite-austenite interfaces in the form of nanoscaled seams (see Figs. 8 and 11). Such a topology might act as a soft barrier against incoming cracks or stress-strain localizations from the martensite. We hence speculate that the austenite seam is a compliance or respectively repair layer that can immobilize defects through its high formability and prevent cracks from percolating from one martensite grain into another (Fig. 11). In this context it is important to note that conventional martensite-martensite interfaces often have a small-angle grain boundary between them which facilitates crack penetration from one lath to another. Here, a compliant austenite seam between the laths might hence be very efficient in stopping cracks. We emphasize this point since the increase in macroscopic ductility can generally be promoted by both, an increase in strain hardening at the later stages of deformation and by mechanisms that prevent premature damage initiation.

## 5. Conclusions

We studied carbon partitioning, retained austenite, austenite stabilization, nanoscale austenite reversion, carbide formation, and kinetic freezing of carbon during heat treatment of a martensitic stainless steel Fe-13.6Cr-0.44C (wt.%). The main results are:

(1) Austenite formation in carbon enriched martensite-austenite interface areas is confirmed by XRD, EBSD, TEM, and APT. Both, the formation of retained austenite and austenite reversion during low-temperature partitioning is discussed. The enrichment of carbon at martensite-martensite grain boundaries and martensite-retained austenite phase boundaries provides the driving force for austenite reversion. The reverted austenite zones have nanoscopic size (about 15-20 nm). The driving forces for austenite reversion are determined by local and not by global chemical equilibrium.

(2) Martensite-to-austenite reversion proceeds fast. This applies to both, the formation of reversed austenite at retained austenite layers and austenite reversion among martensite laths. The volume fraction of austenite has nearly doubled after 2 minutes at 400°C.

(3) The carbides formed during tempering have $M_3C$ structure. With increasing tempering time the dispersion of the carbides decreases due to the Gibbs-Thomson effect.

(4) During tempering between 300°C and 500°C carbon redistributes in three different ways: During quenching, in the vicinity of martensite-austenite interfaces, carbon segregates from the supersaturated martensite to both, the hetero-interfaces and to homophase grain boundaries. During tempering, carbon continuously partitions to martensite-austenite interfaces, driving the carbon enriched areas towards austenite reversion (irrespective of whether the nucleation zones were initially retained or reversed austenite). Carbon inside martensite, far away from any interfaces, tends to form $M_3C$ carbides. This means that carbon segregates to martensite grain boundaries, to martensite-austenite interfaces, and forms carbides.

(5) We differentiate between 3 different types of austenite, namely, first, as-quenched retained austenite with nominal carbon content and low stability; second, retained austenite with increased carbon content and higher stability due to partitioning according to the local chemical potential of carbon; and third, reverted austenite.

(6) The nanoscale structural changes lead to drastic improvements in the mechanical properties. A sample after 1 minute tempering at 400°C has 2 GPa tensile strength with 14% total elongation. The strength increase is attributed to the high carbon content of the martensite and the interaction between dislocations and nano-sized carbides. The TRIP effect of the austenite during deformation, including the reverted nano-scale austenite, contributes to a strain hardening capacity and, hence, promotes the ductility. Also, the topology of the reverted austenite is important: We suggest that the nanoscaled seam topology of the austenite surrounding the martensite acts as a soft barrier which has compliance and repair function. This might immobilize defects and prevent cracks from growth and inter-grain percolation.

(7) We attribute the fast nanoscale austenite reversion to an effect that we refer to as kinetic freezing of carbon. This means that the carbon is fast inside the martensite when leaving it but slow (and, hence, frozen) when entering the austenite. This means that carbon becomes trapped and highly enriched at the martensite-austenite interfaces owing to its low mobility within the austenite during low-temperature partitioning. This provides a much higher local driving force for austenite reversion. This means that the formation of nanoscaled reverted austenite depends mainly on the local but not on the global chemical potential of carbon at internal interfaces.

## Acknowledgements

The authors are grateful for discussions on carbon partitioning and martensite tempering with Professor George D.W. Smith from Oxford University.

## Tables

Table 1: Chemical composition of material used for the investigation (1.4034, X44Cr13, AISI 420).

| | C | Cr | Mn | Ni | Si | N | Fe |
|---|---|---|---|---|---|---|---|
| wt.% | 0.437 | 13.6 | 0.53 | 0.16 | 0.284 | 0.0205 | Bal. |
| at.% | 1.97 | 14.19 | 0.52 | 0.15 | 0.55 | 0.079 | Bal. |

Table 2. Change of the carbon content observed in each phase via atom probe tomography during annealing, quenching, and austenite reversion. The carbon partitioning to the different phases is quantified in terms of the enrichment factor ε=(at.% C tempered)/(at.% C as quenched) which allows us to compare the chemical composition in the phases before and after tempering.

| State of samples | Retained austenite (at.%) | martensite (at.%) | interface (at.%) | reverted austenite (at.%) |
|---|---|---|---|---|
| nominal composition (annealing at 1150°C) | 1.97% | -- | -- | -- |
| as quenched | 1.90% | 0.98% | 4.52% | -- |
| quenched plus tempering (400°C/1 min) | -- | 0.82% (ε=0.84) | -- | 6.86% (ε=3.61) |
| quenched plus tempering (400°C/30 min) | 2.42% (ε=1.27) | 0.48% (ε=0.49) | -- | -- |

Table 3. Diffusion data for carbon in ferrite and austenite taken from [40]. For the current heat treatment case of 400°C (673K) the diffusion coefficient in ferrite is $D_\alpha = 2\times10^{-12}$ $m^2/s$ and in austenite $D_\gamma = 5\times10^{-17}$ $m^2/s$. The table gives the mean free path for the different tempering stages. The diffusion of carbon on the ferrite can be regarded as a lower bound. The corresponding value for martensite is likely to be higher owing to the high defect density of the martensite.

| Time / min | Austenite / m | Ferrite / m |
|---|---|---|
| 1 | $1.3\times10^{-7}$ | $2.7\times10^{-5}$ |
| 2 | $1.9\times10^{-7}$ | $3.8\times10^{-5}$ |
| 30 | $7.4\times10^{-7}$ | $1.5\times10^{-4}$ |

**Figure(s)**

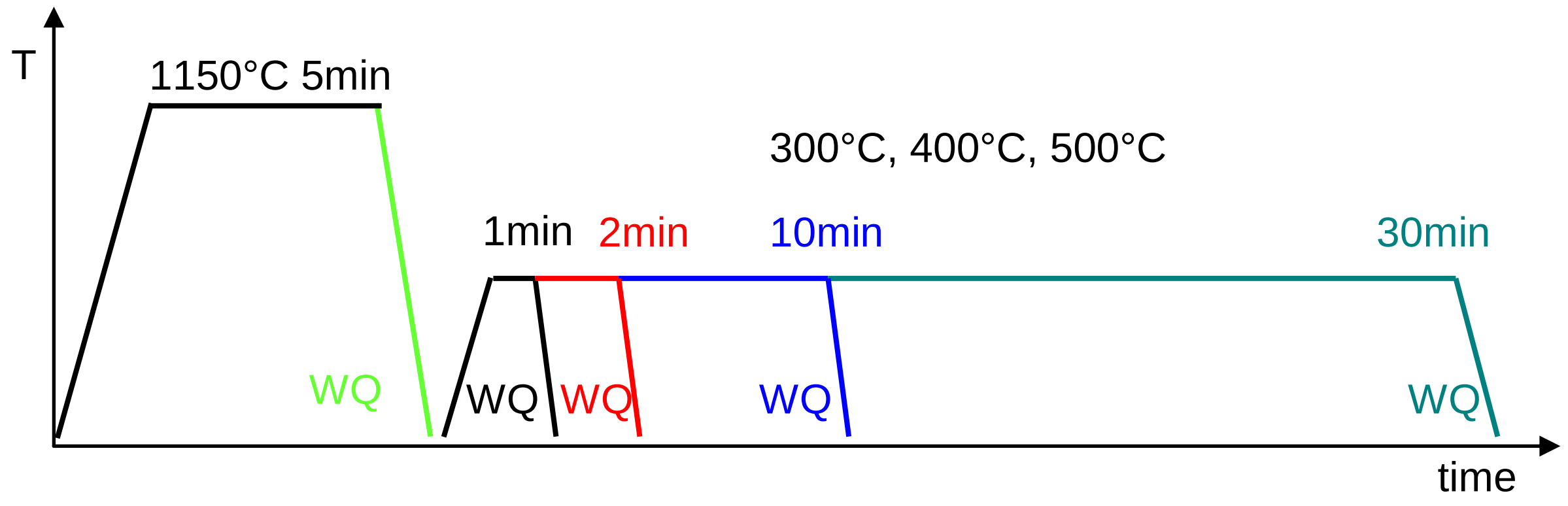

Fig. 1 Schematic diagram of the heat treatment route (WQ: water quenching).

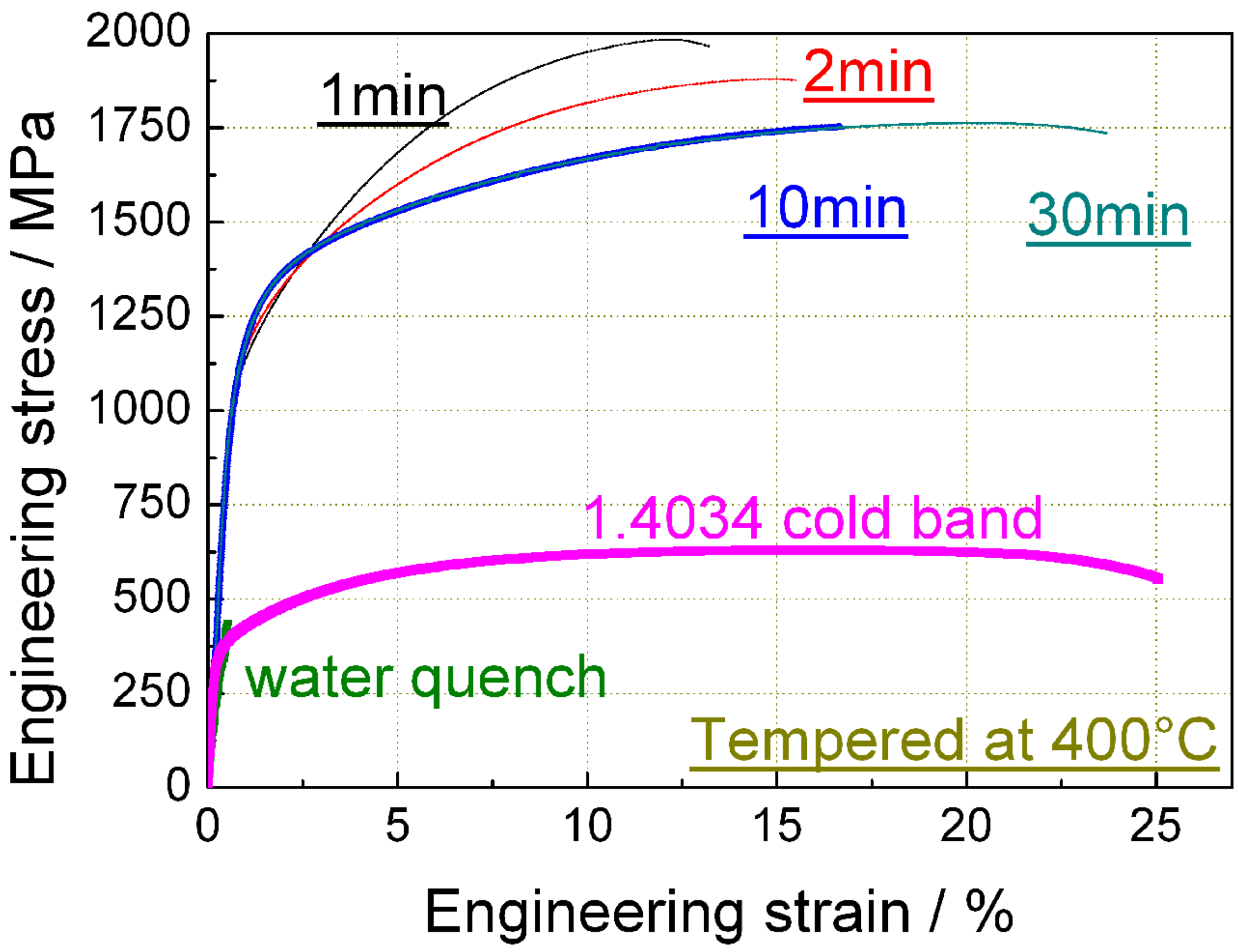

(a)

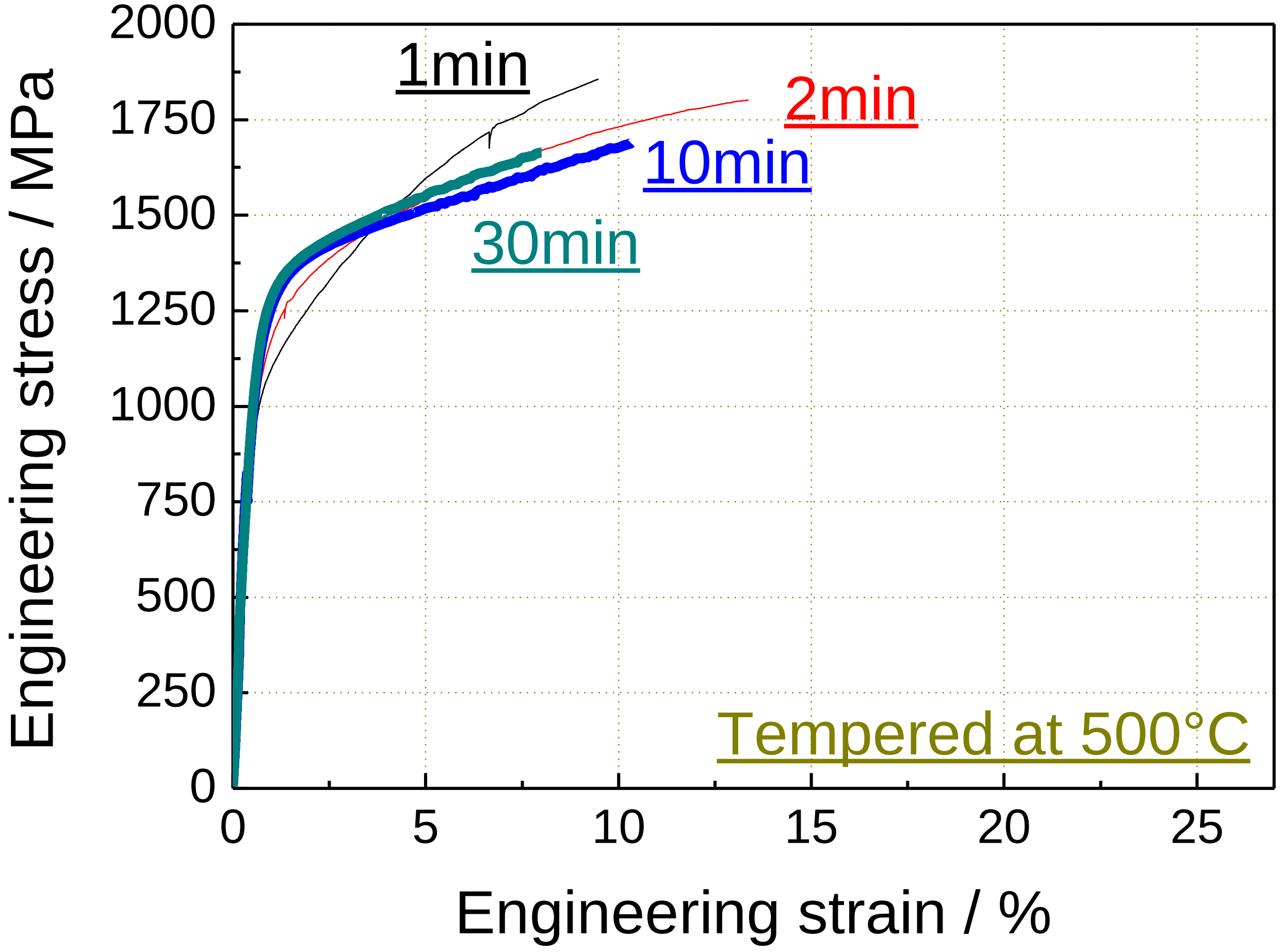

(b)

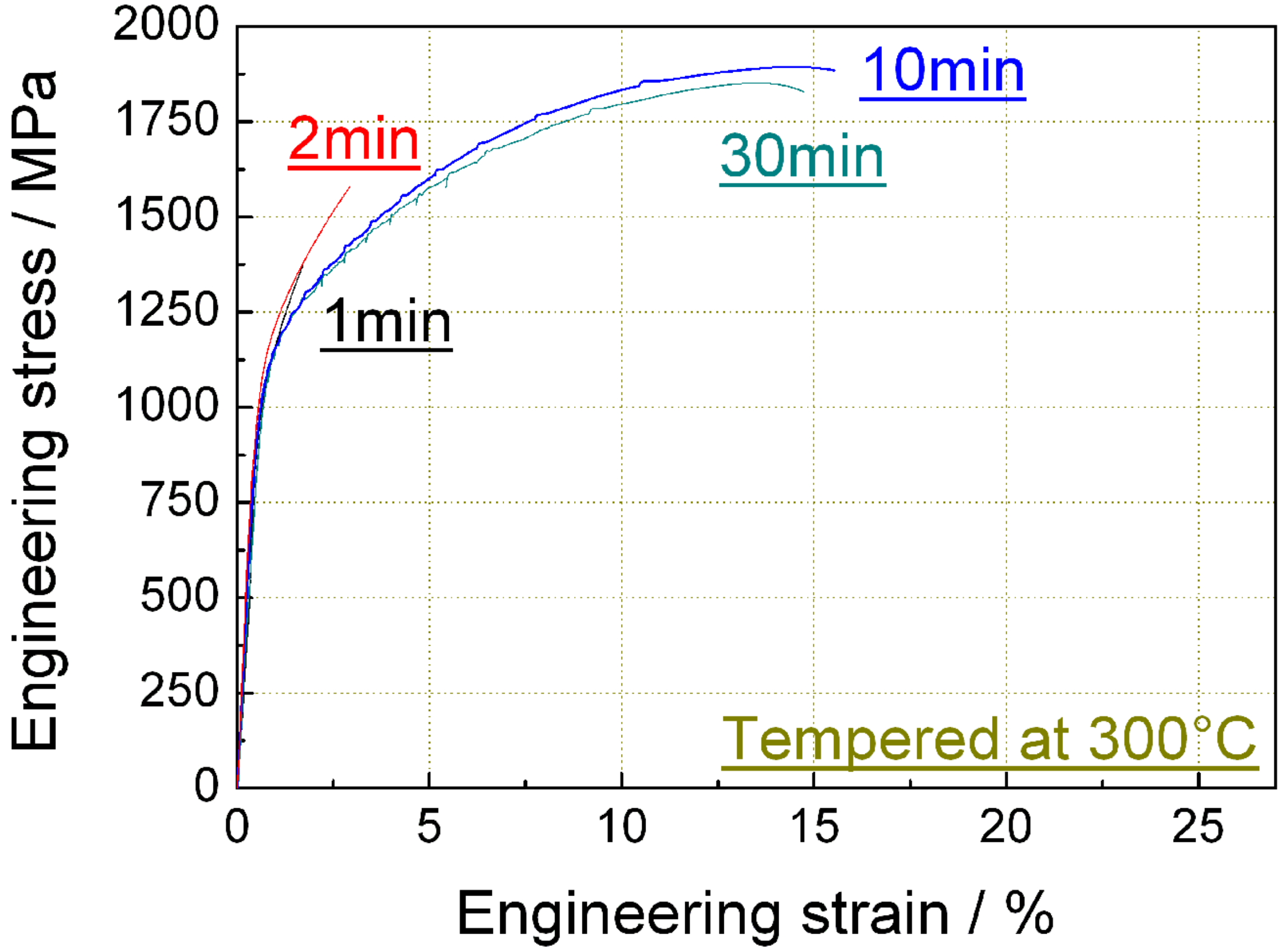

(c)

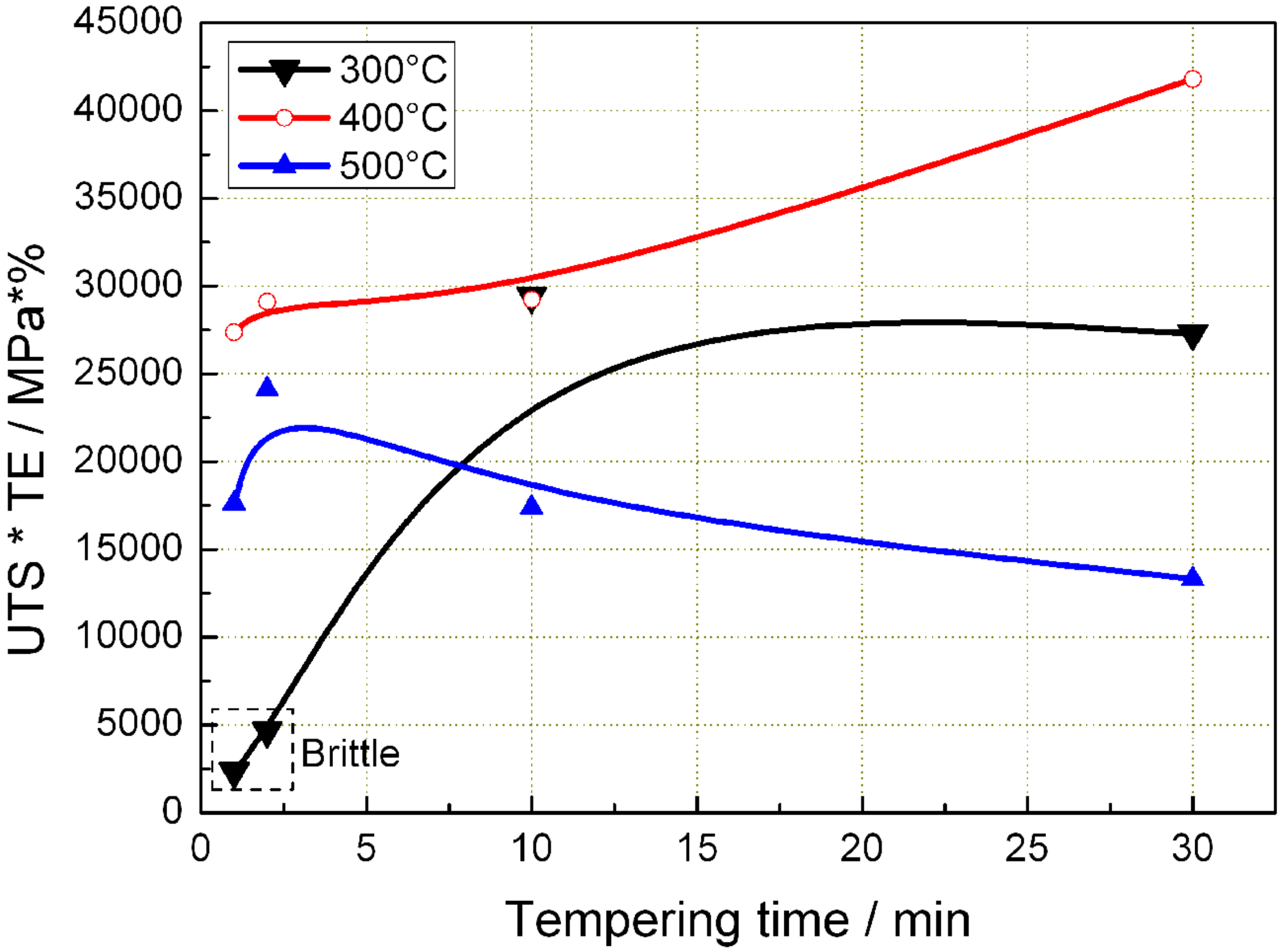

(d)

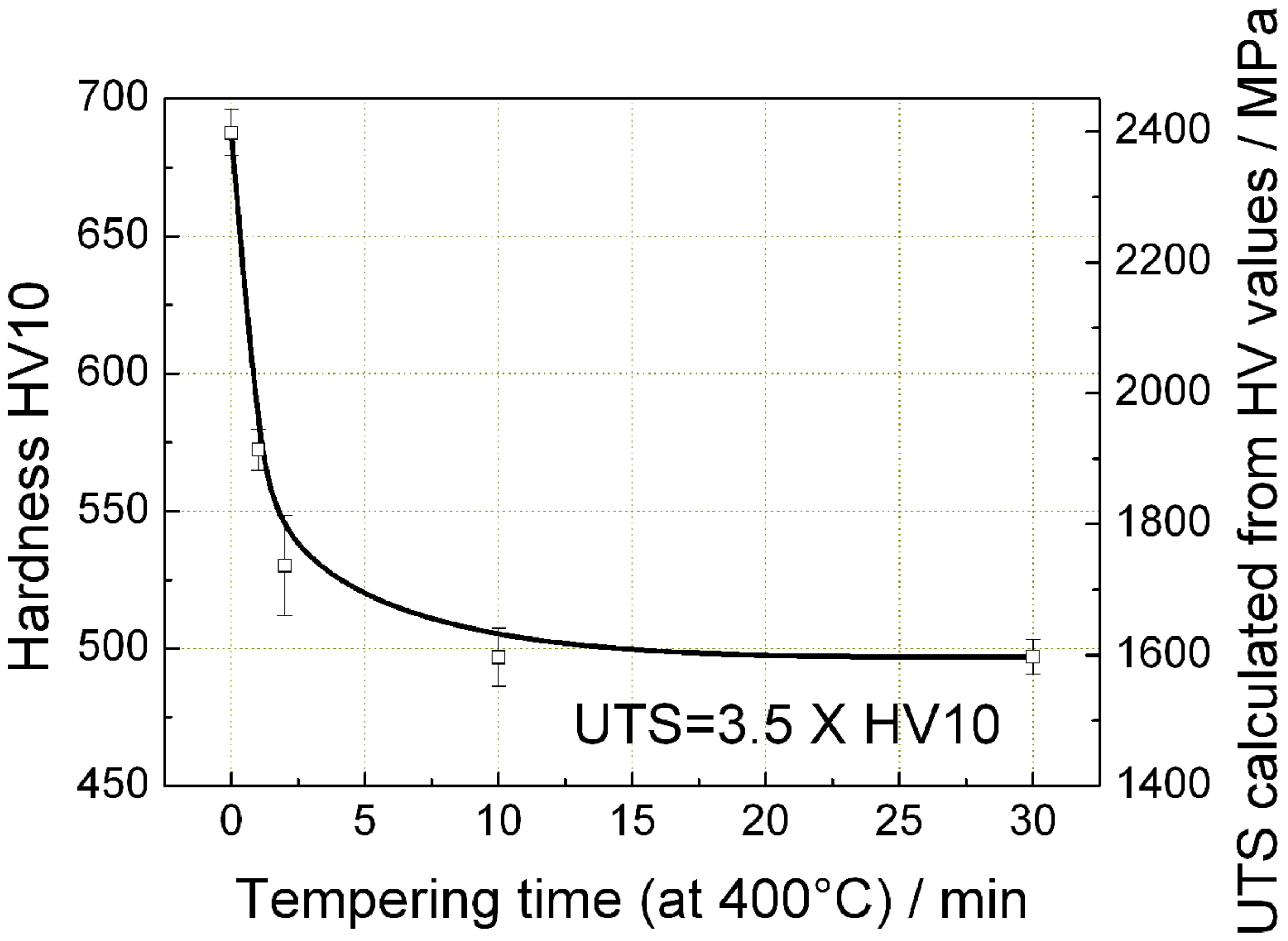

(e)

Fig. 2 Mechanical properties of the quenched and partitioned stainless steel Fe-13.6Cr-0.44C (wt.%, 1.4034, AISI 420) after different types of partitioning and austenite reversion treatments. The original state of a commercial alloy (1.4034, cold band) is shown as reference. The term 'cold band' refers here to hot rolled, cold rolled, and finally recrystallized material.
(a-c) stress strain curve of samples tempered at 400°C, 500°C, 300°C, respectively. Note in (a) that the as-quenched sample (green) fails already in the elastic regime; (d) Multiplied quantity UTS ×TE as a function of annealing time for the three different temperatures; (e) UTS-HV relationship. (UTS: ultimate tensile strength; HV: Vickers hardness; TE: total elongation)

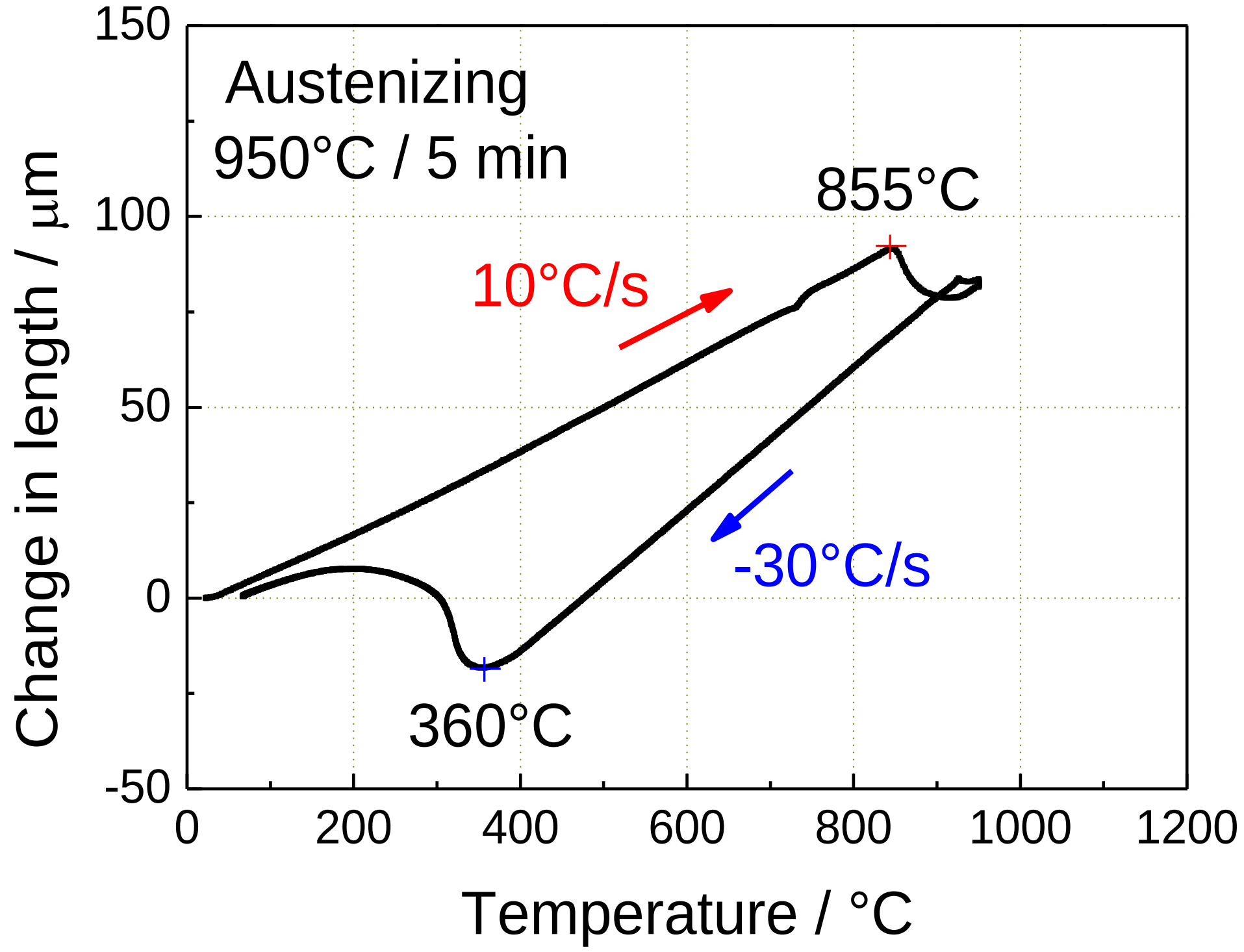

(a)

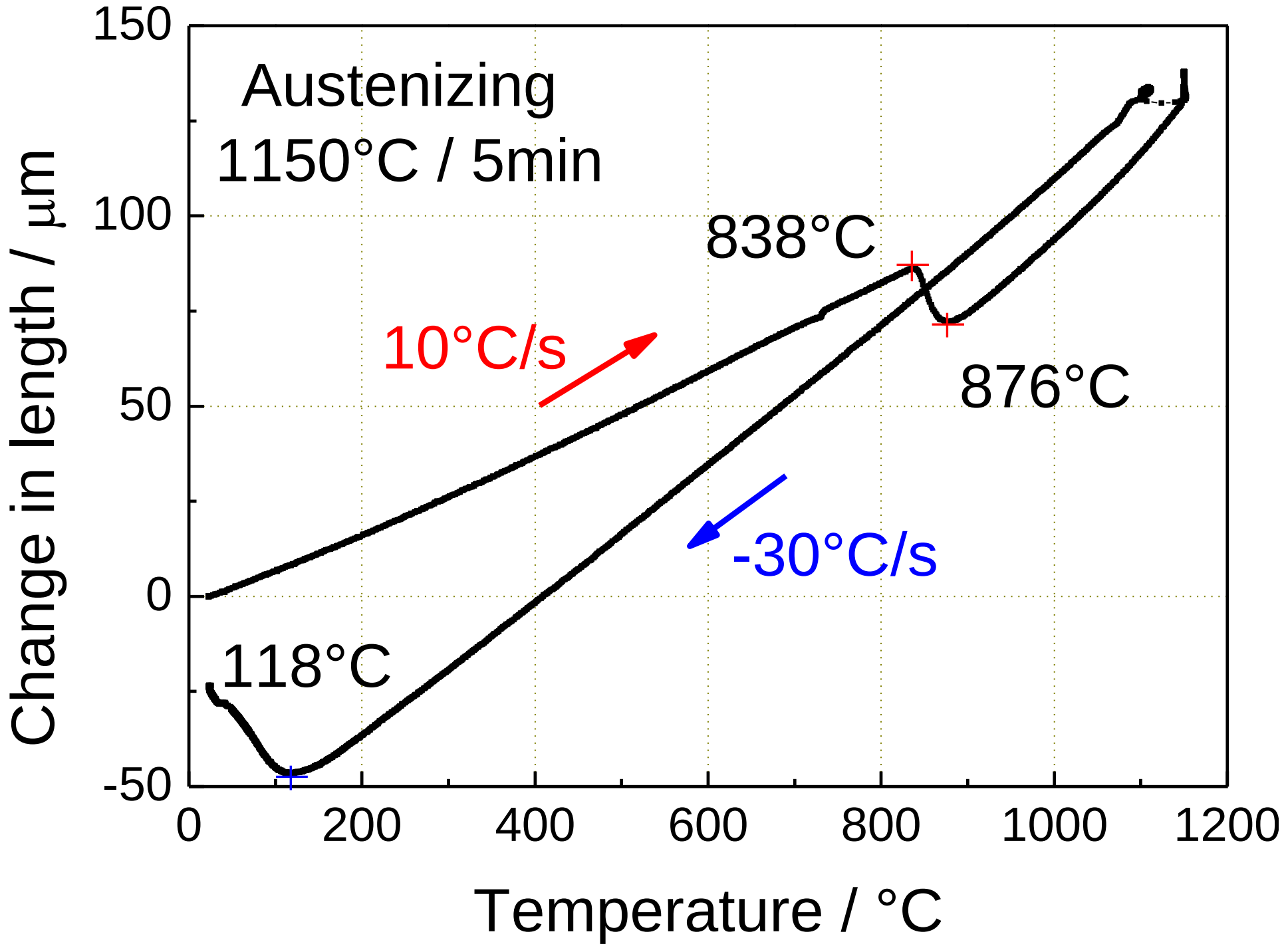

(b)

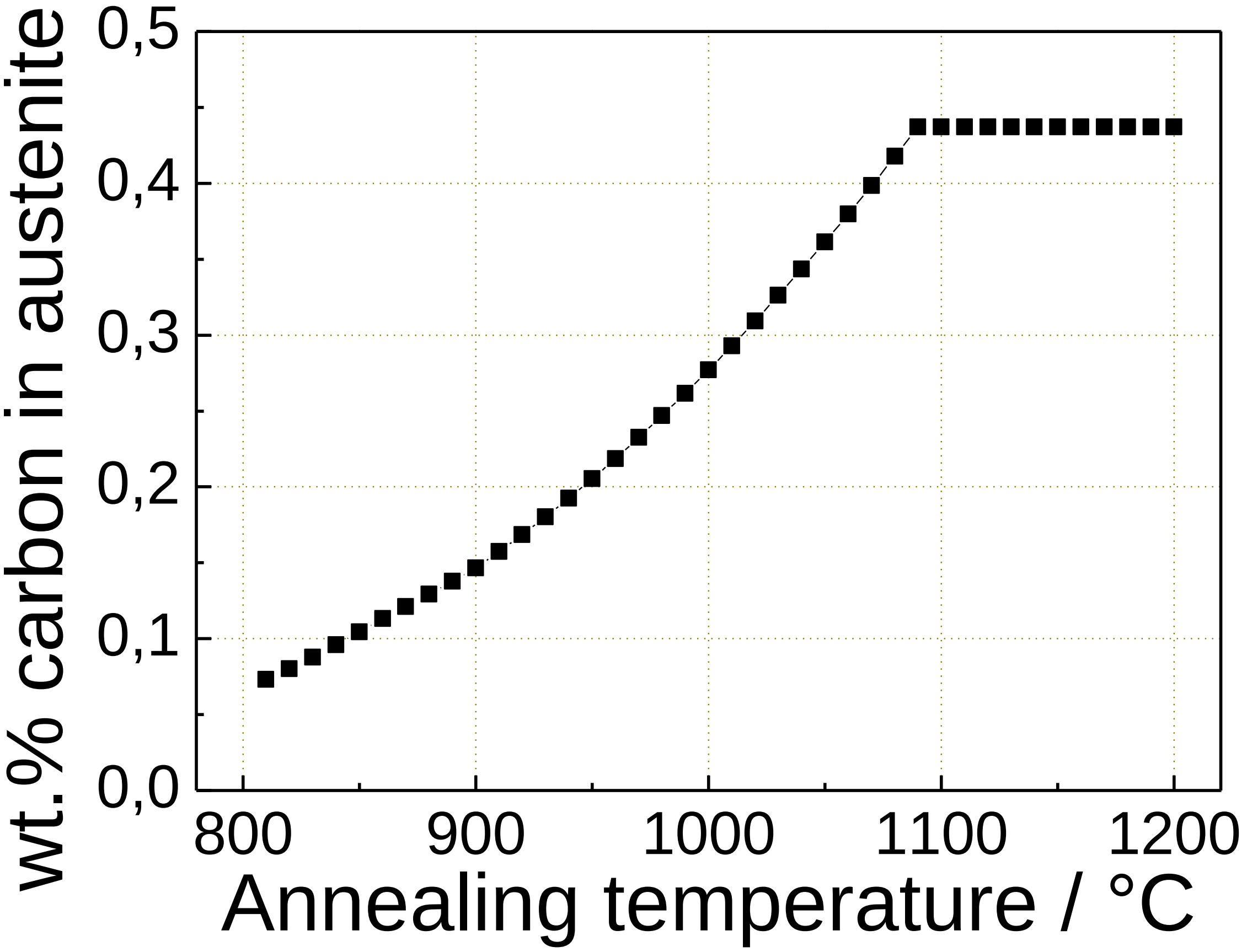

(c)

Fig. 3 Results of the dilatometer tests of the stainless steel Fe-13.6Cr-0.44C (wt.%, 1.4034).
(a) Austenitization at 950°C for 5min (b) Austenitization at 1150°C for 5min (c) calculated equilibrium carbon content in austenite at different annealing temperature (Thermo-Calc TCFE5)

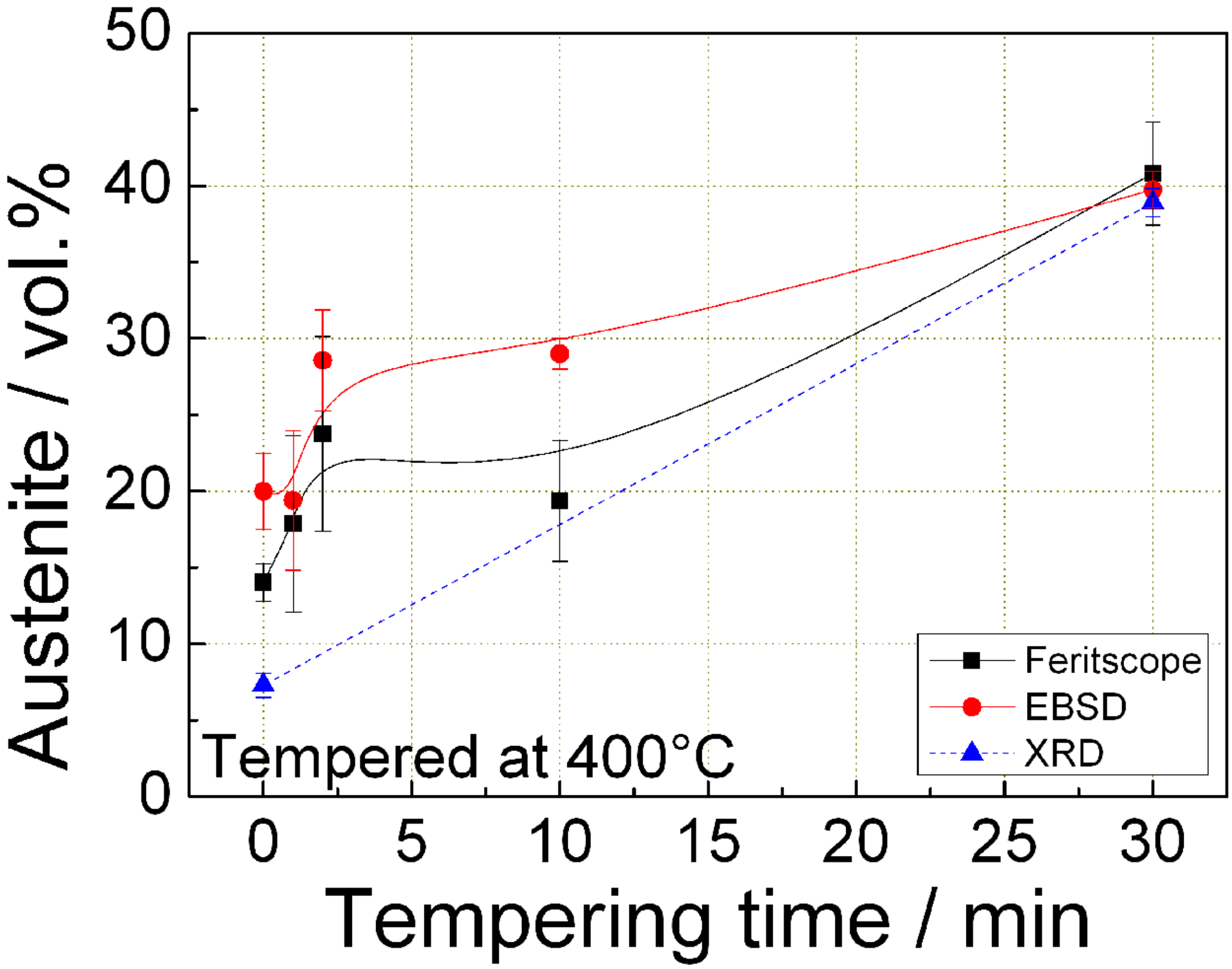

Fig. 4 Austenite volume fraction as a function of tempering time (at 400°C) measured by feritscope (magnetic signal), EBSD, and XRD.

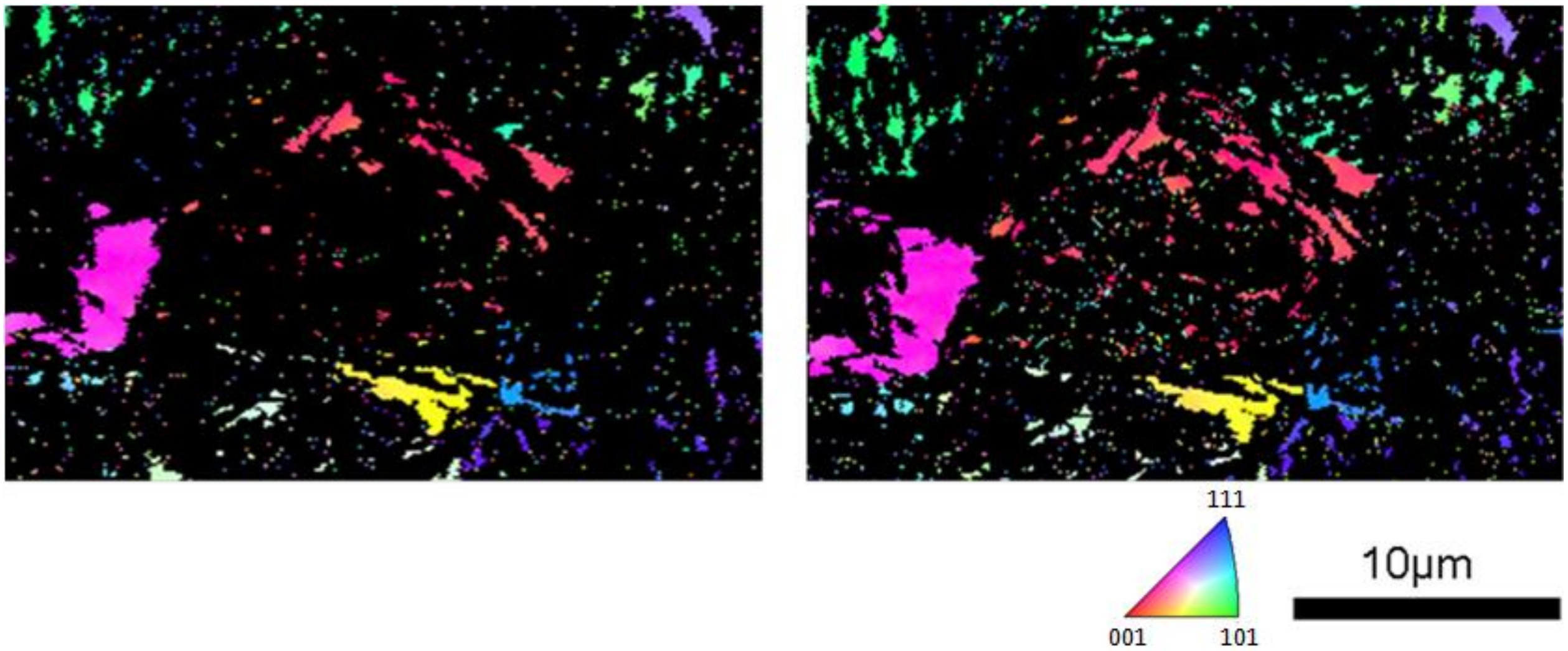

(a) (b)

Fig. 5 EBSD inverse pole figure map of the same specimen region showing retained and reverted austenite (IPF||ND, only austenite shown): (a) shows the material as quenched containing only retained austenite. (b) shows the material containing both, retained plus reverted austenite after quenching and 5 minutes tempering at 400°C (EBSD: Electron back scattering diffraction; IPF: inverse pole figure color code; ND: normal direction).

Figure(s)

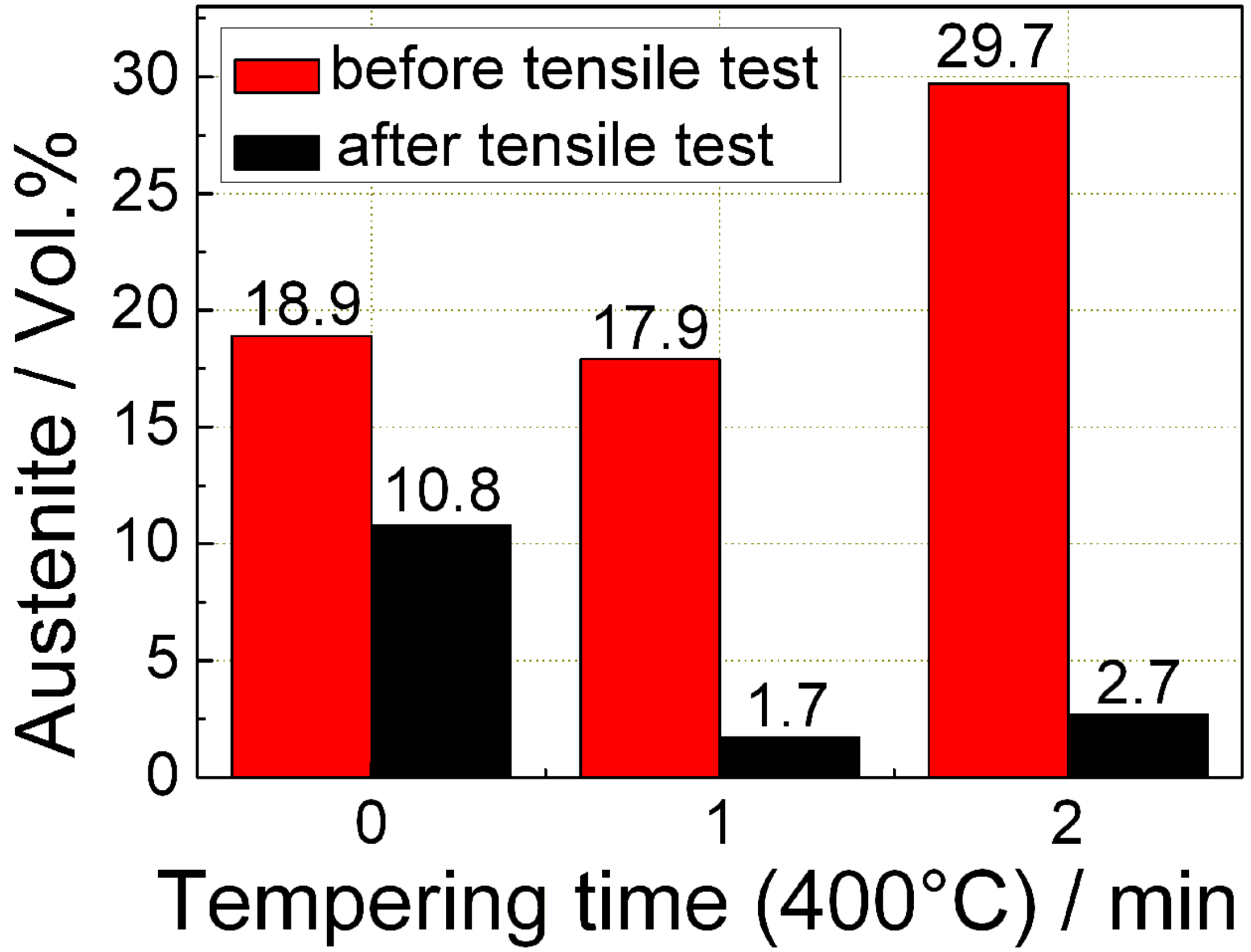

(a)

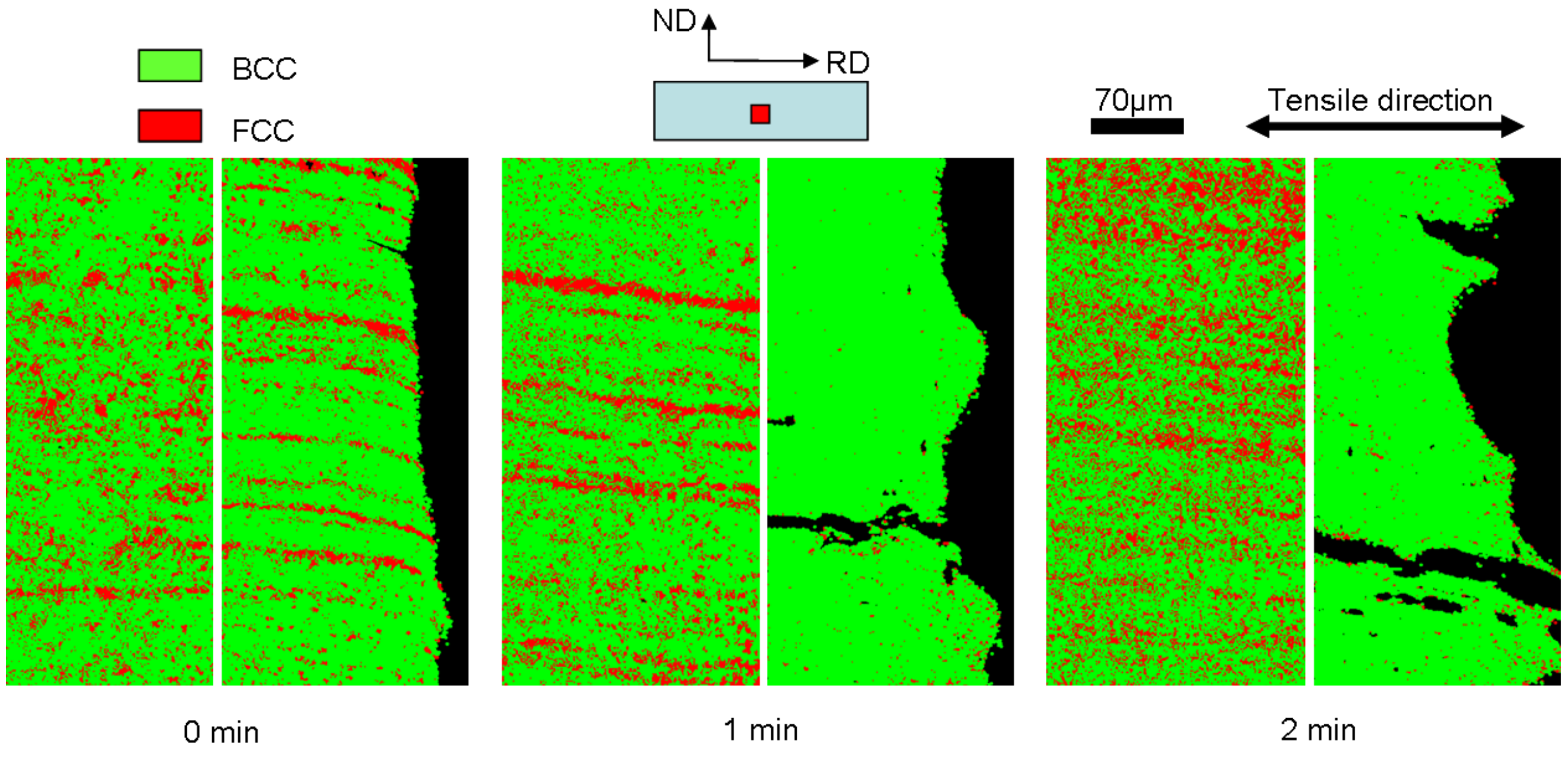

(b)

Fig. 6 (a) EBSD phase maps of samples tempered at 400°C for 0, 1, and 2 minutes at 400°C, and of the TRIP effect obtained from EBSD phase analysis. The red columns show the austenite content before and the black ones after the tensile tests. (b) Microstructure of samples before and after tensile testing subjected to different tempering conditions (left: before tensile test; right: after tensile test); BCC: martensite phase; FCC: austenite phase.

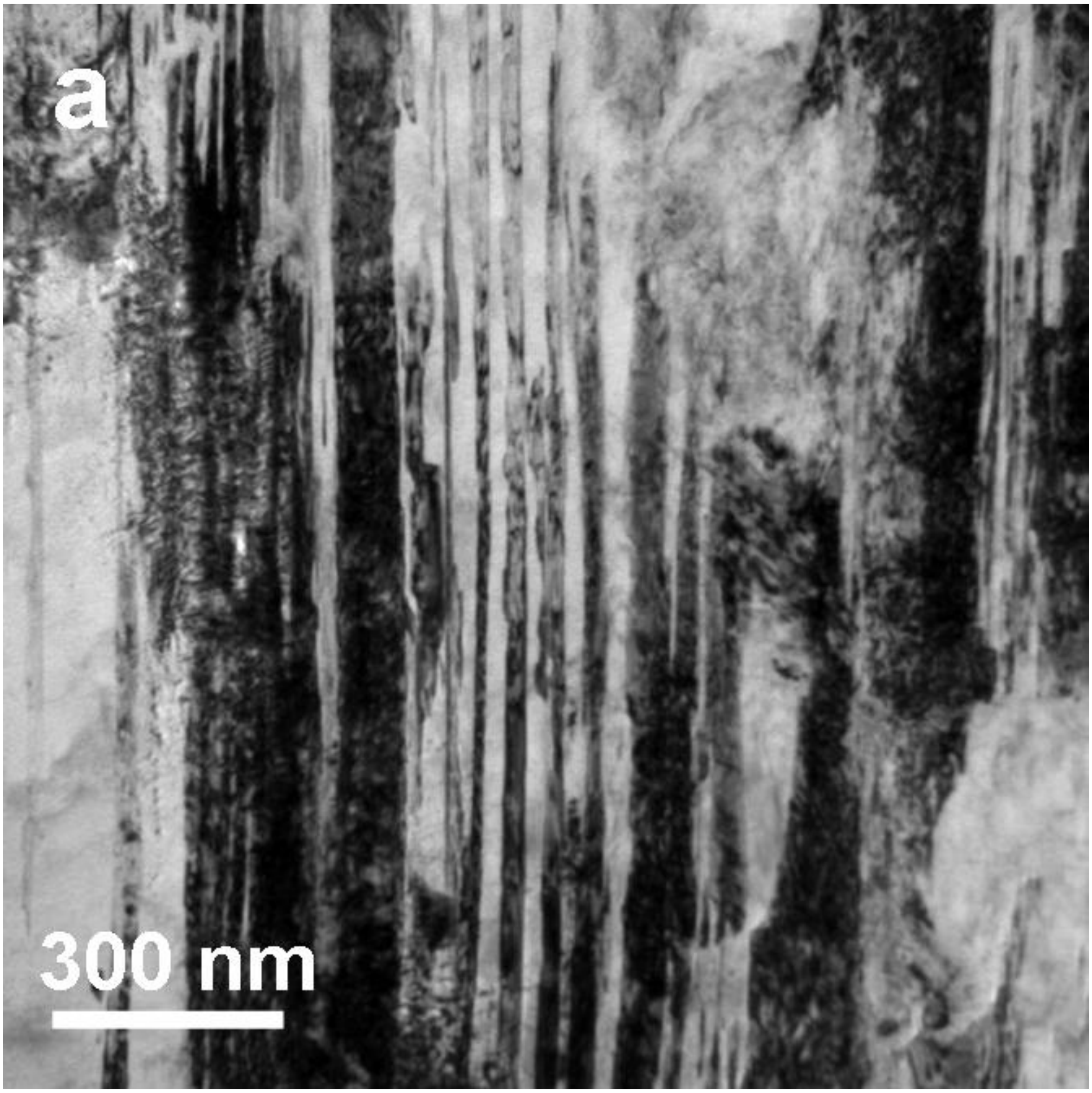

(a)

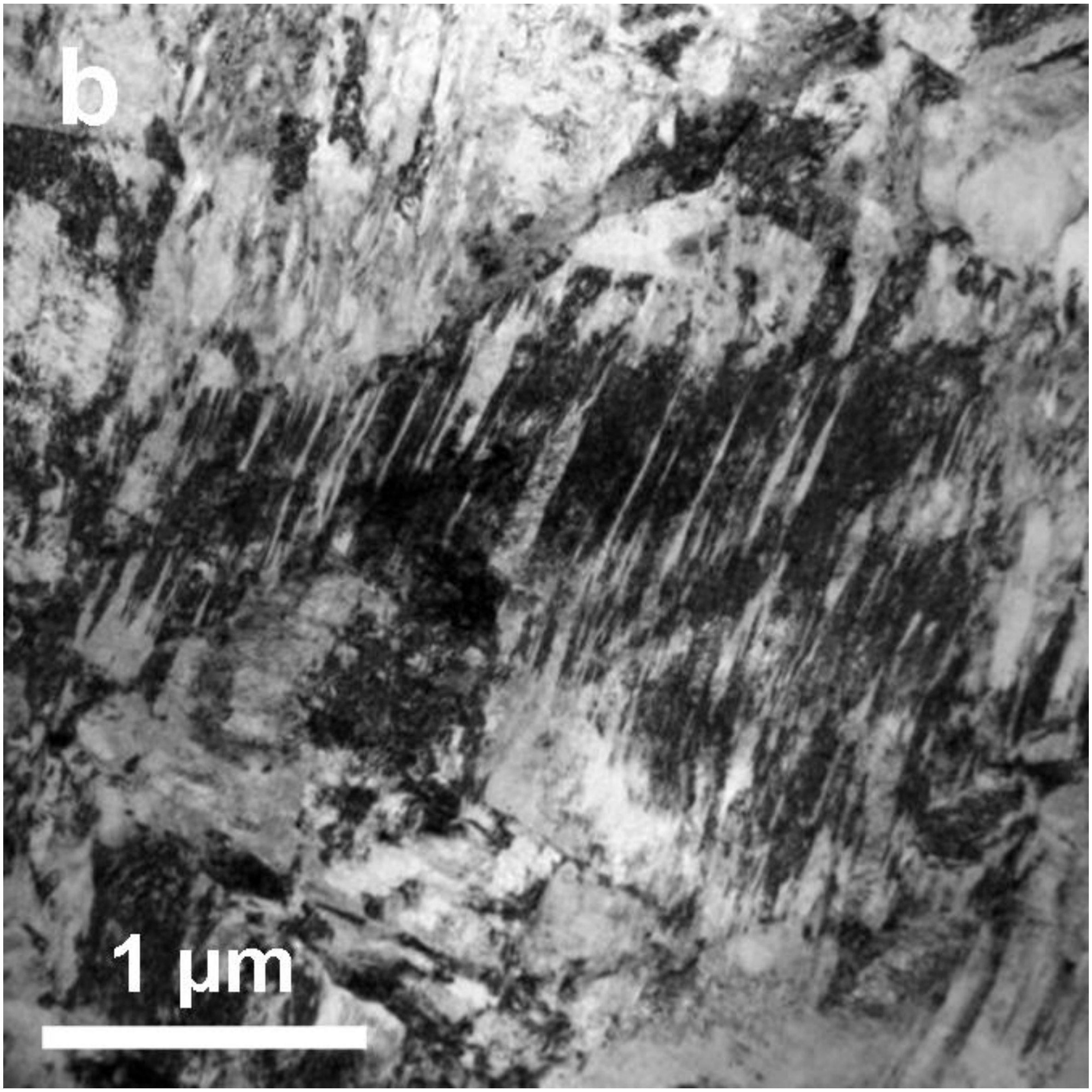

(b)

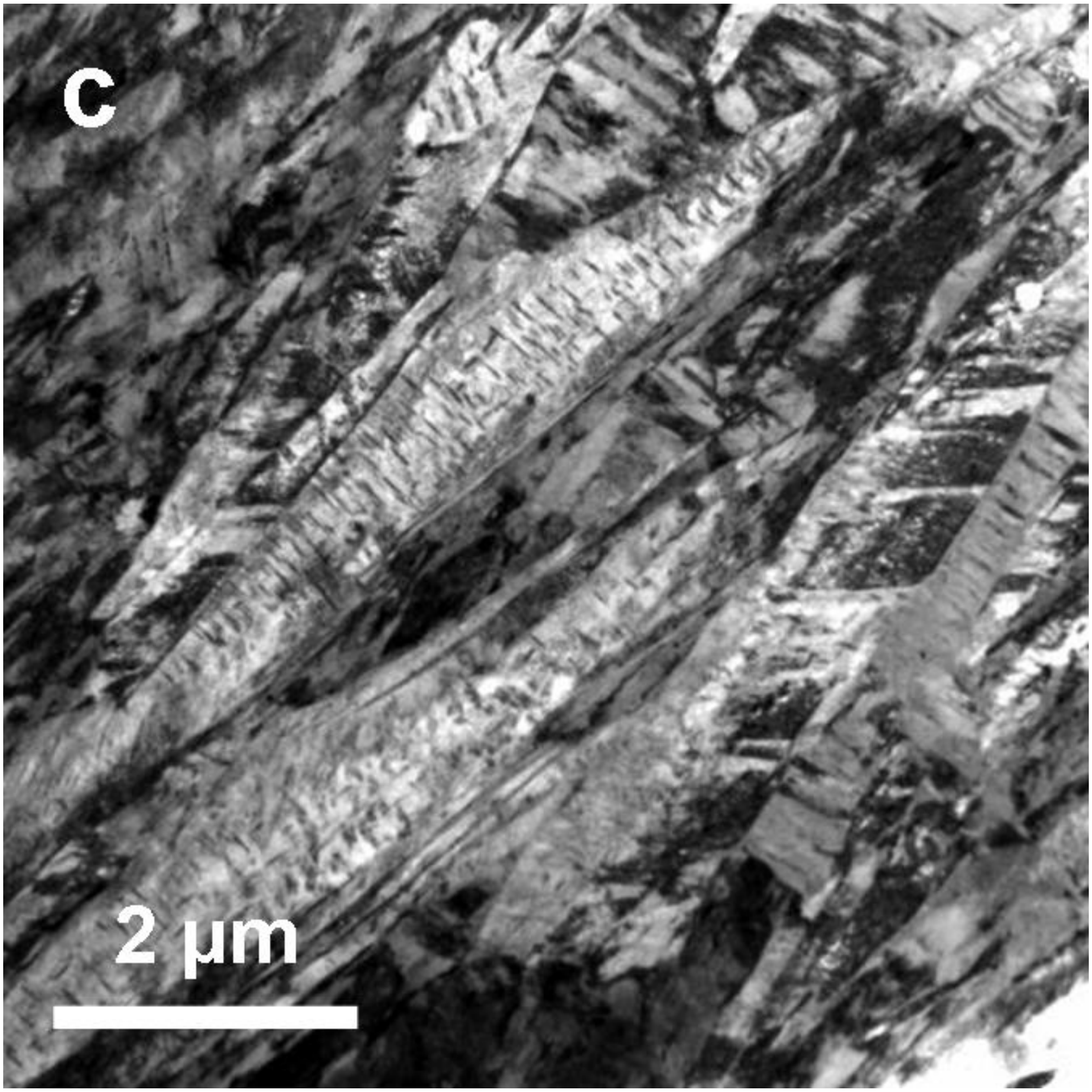

(c)

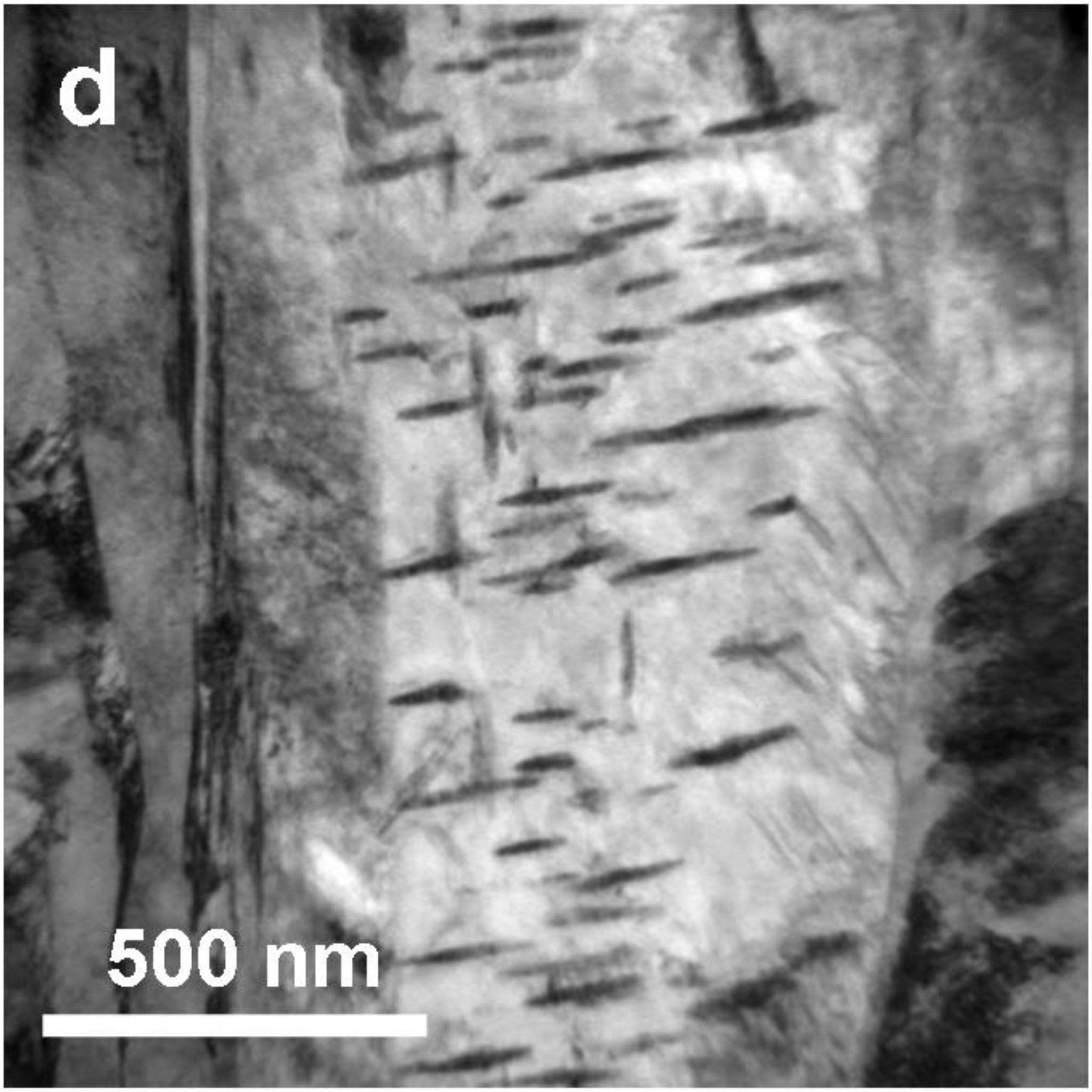

(d)

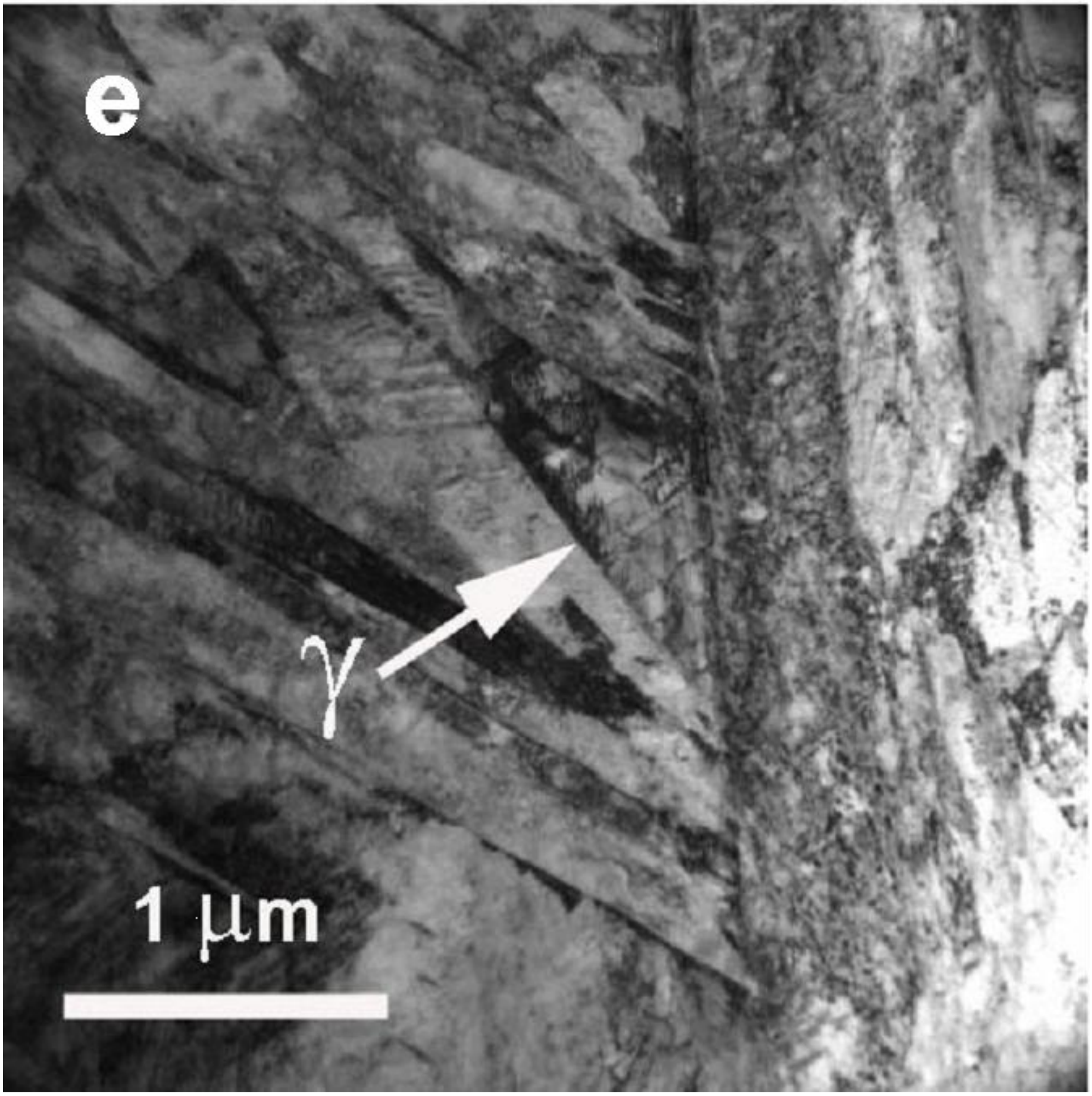

(e)

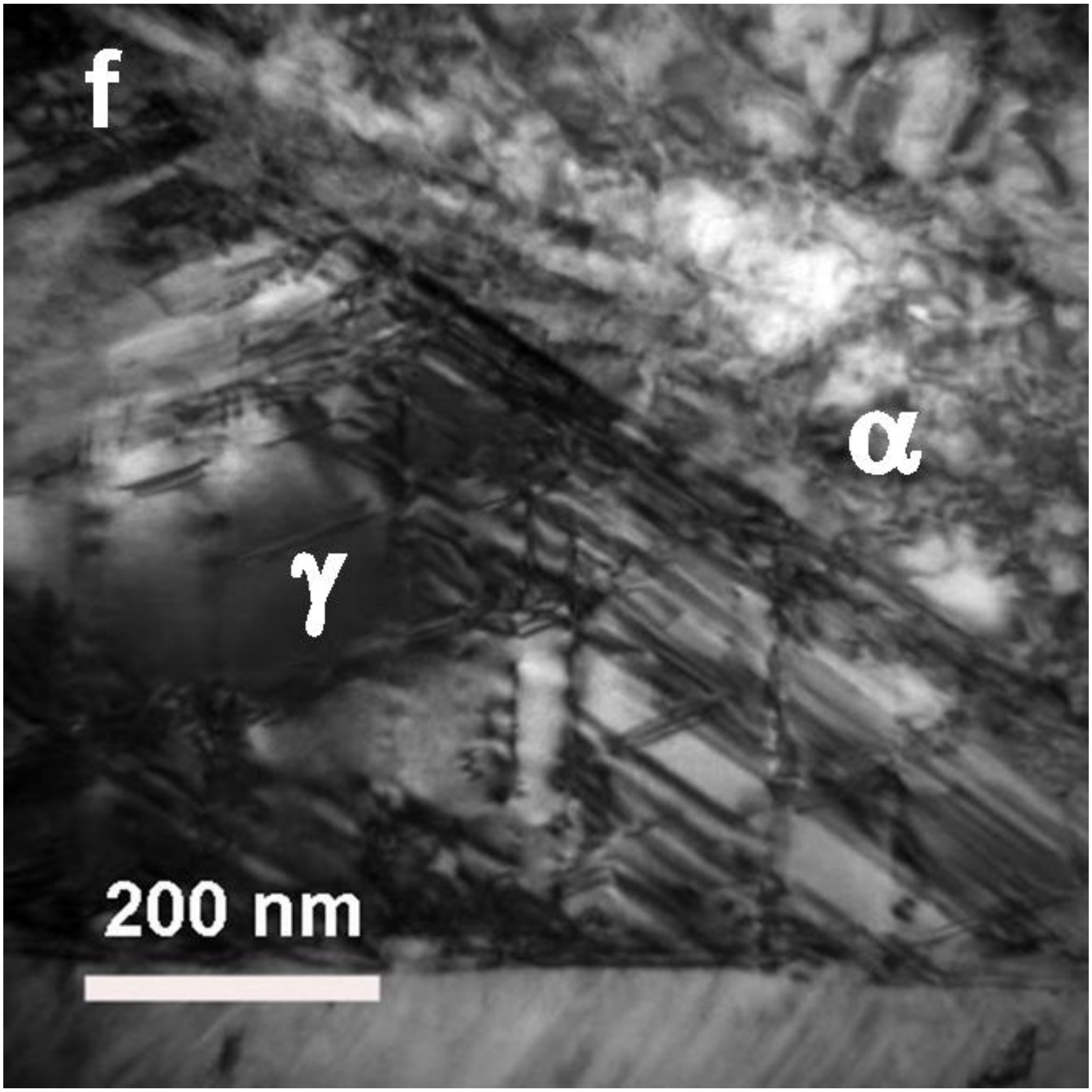

(f)

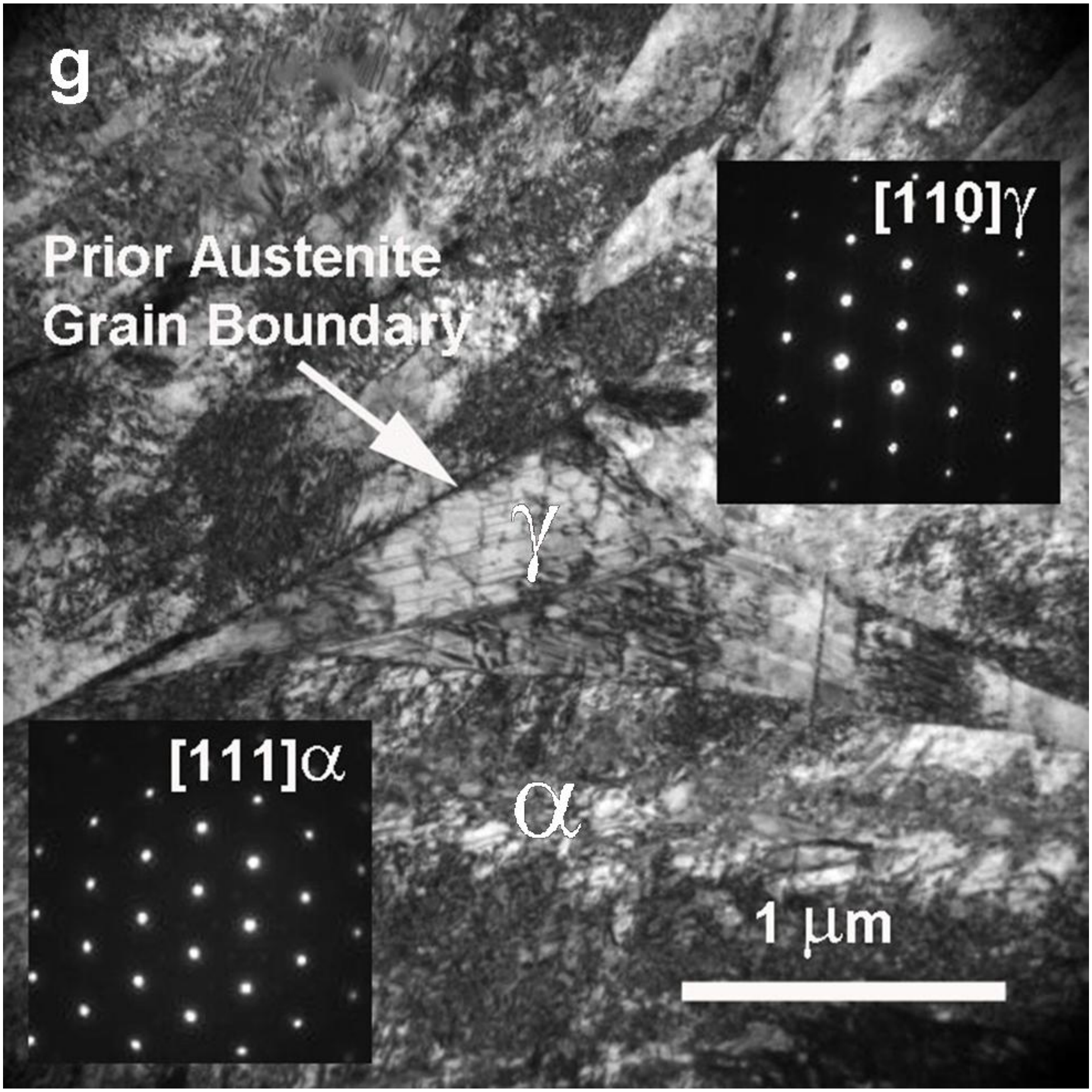

(g)

Fig. 7 TEM images of as quenched sample (only lath martensite was found, a,b);
TEM images of samples tempered at 400°C for 1min: c) Overview image of the very dense array of nanoscaled carbides that is formed during tempering; d) In-grain view of the carbides; e) Overview image of the formation of a reverted austenite grain that is located at a former martensite-martensite grain boundary; f) Close-up view of reverted austenite that is surrounded by martensite; g) Electron diffraction analysis reveals that a Kurdjumov-Sachs growth orientation relationship exists between the martensite matrix and the reverted austenite.

Figure(s)

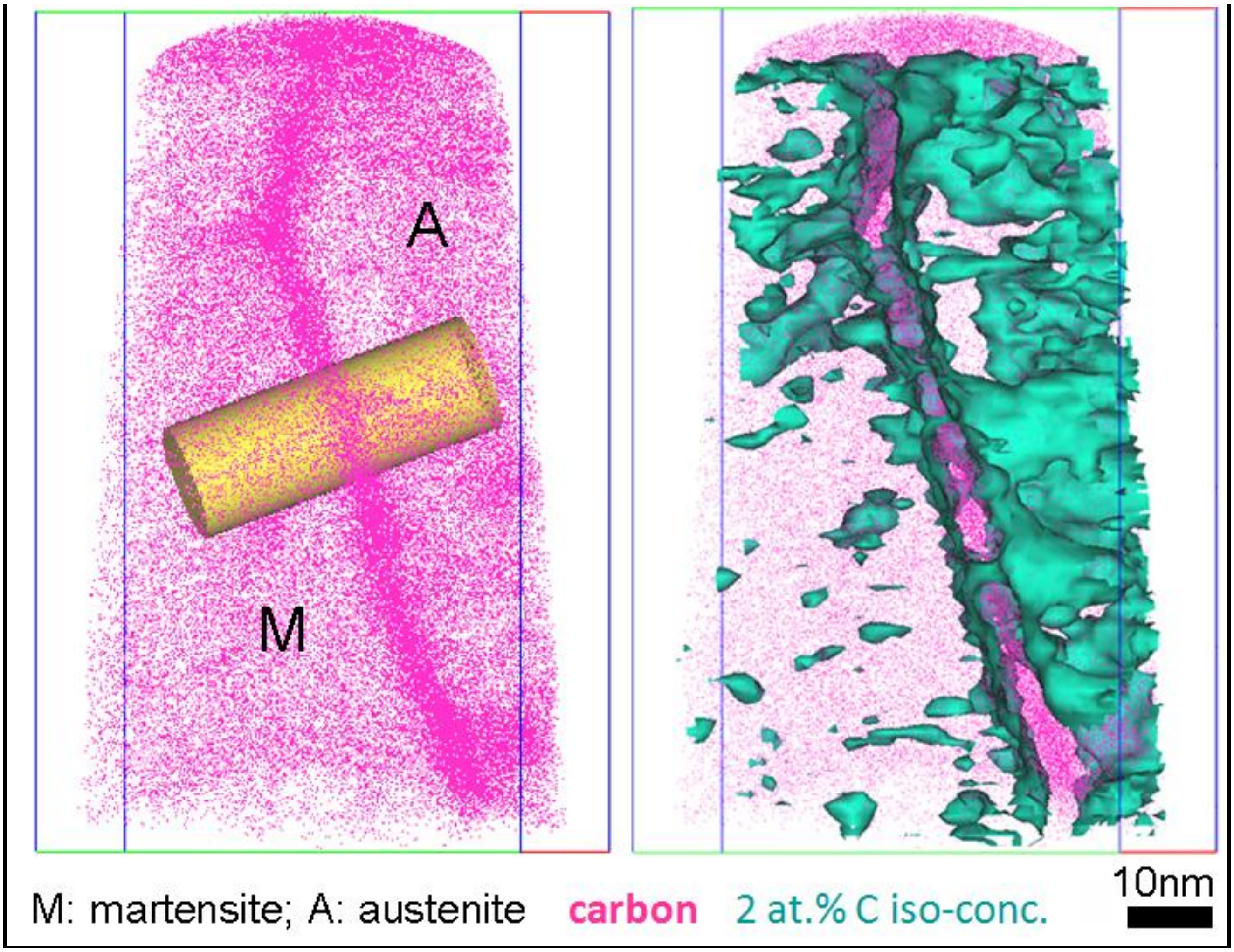

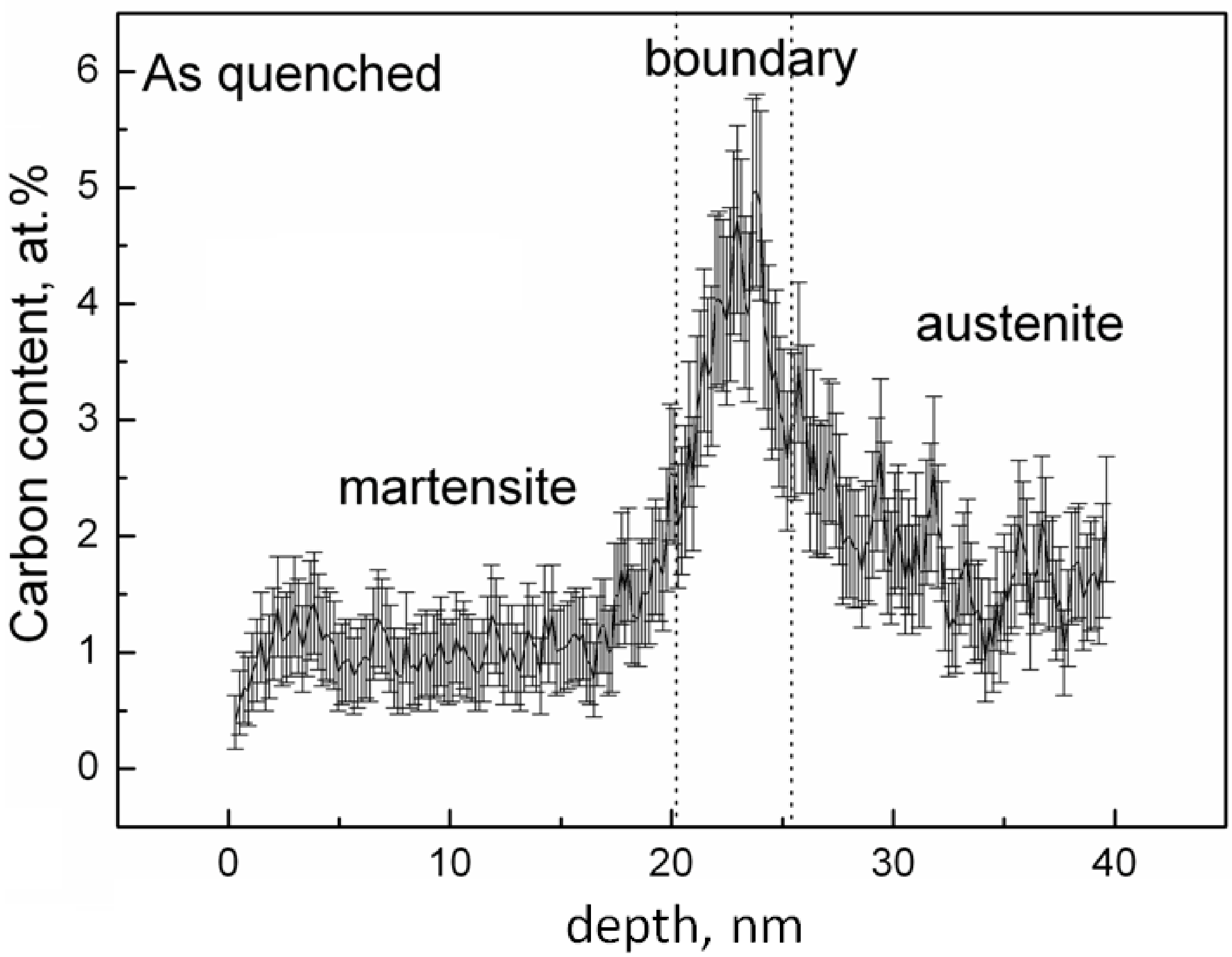

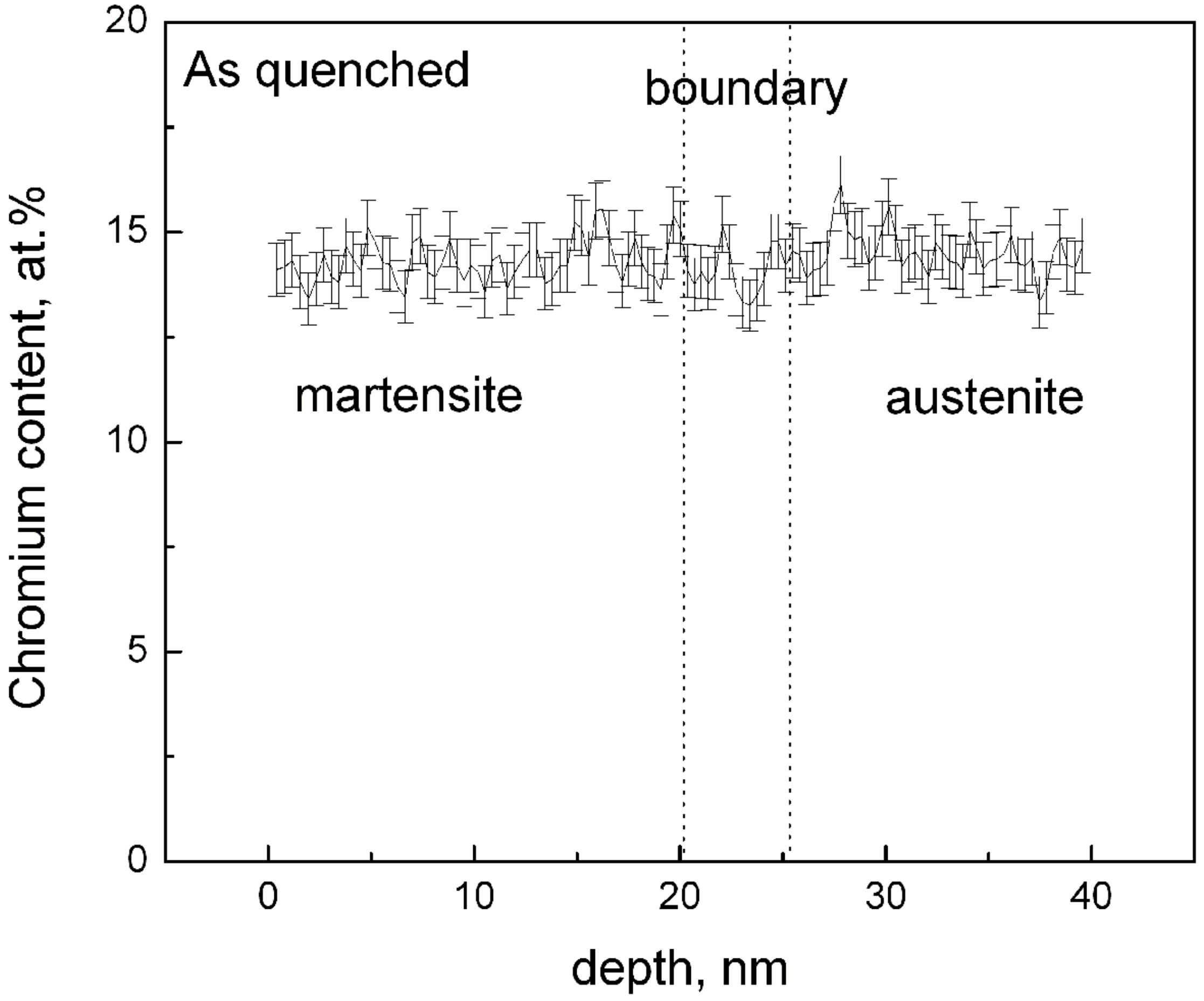

(a)

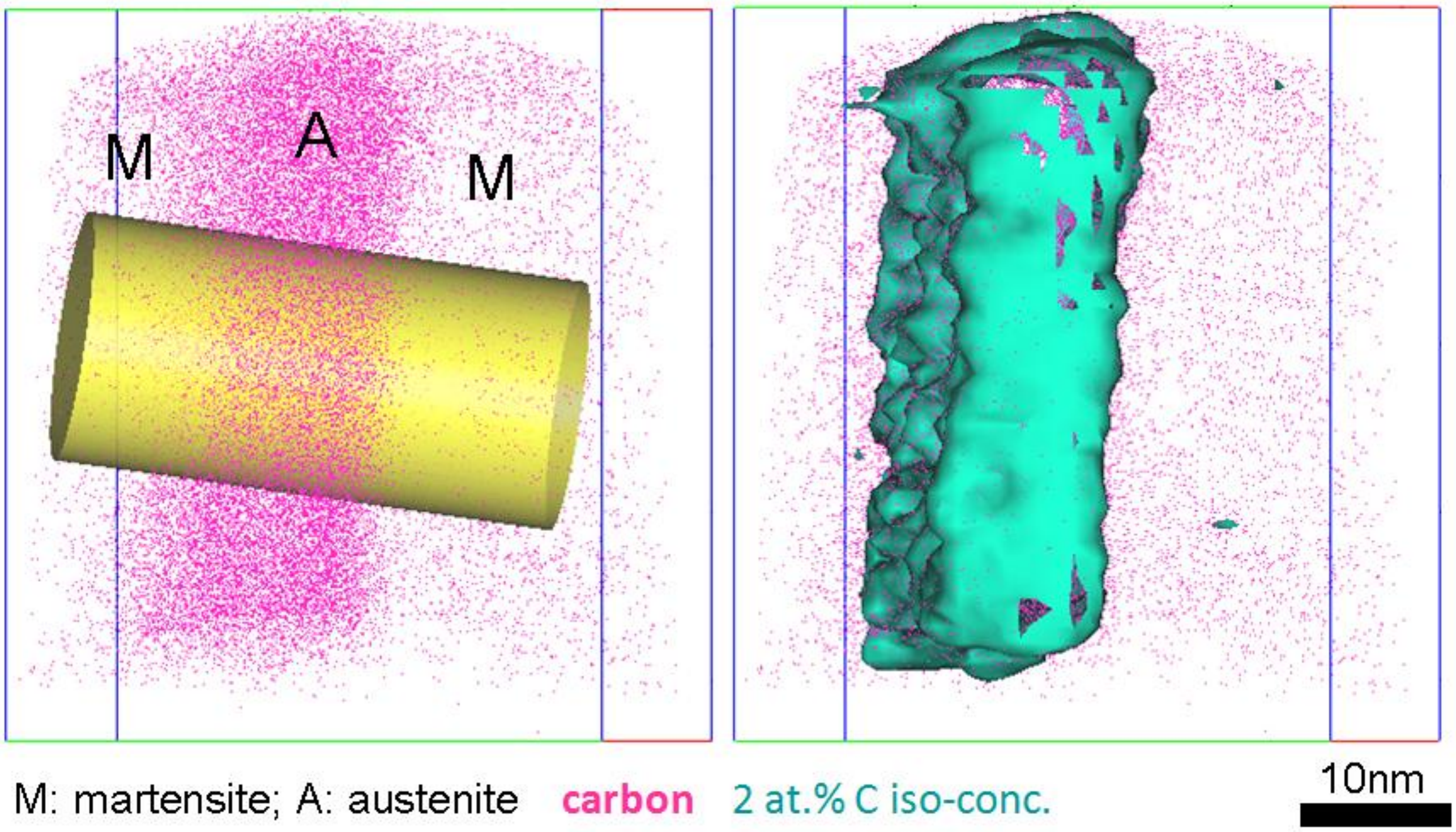
M
A
M
M: martensite; A: austenite
carbon
2 at.% C iso-conc.
10nm

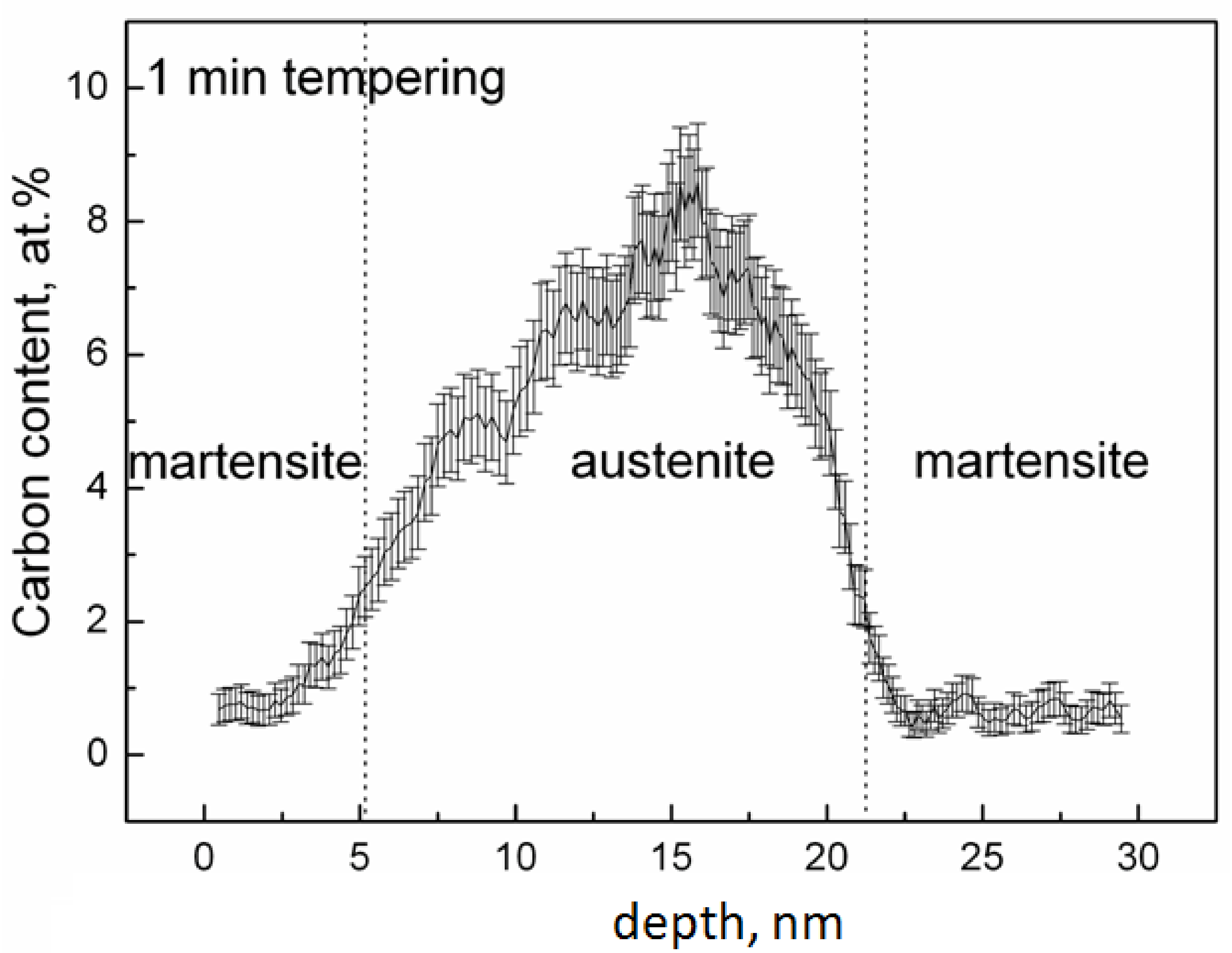

(b)

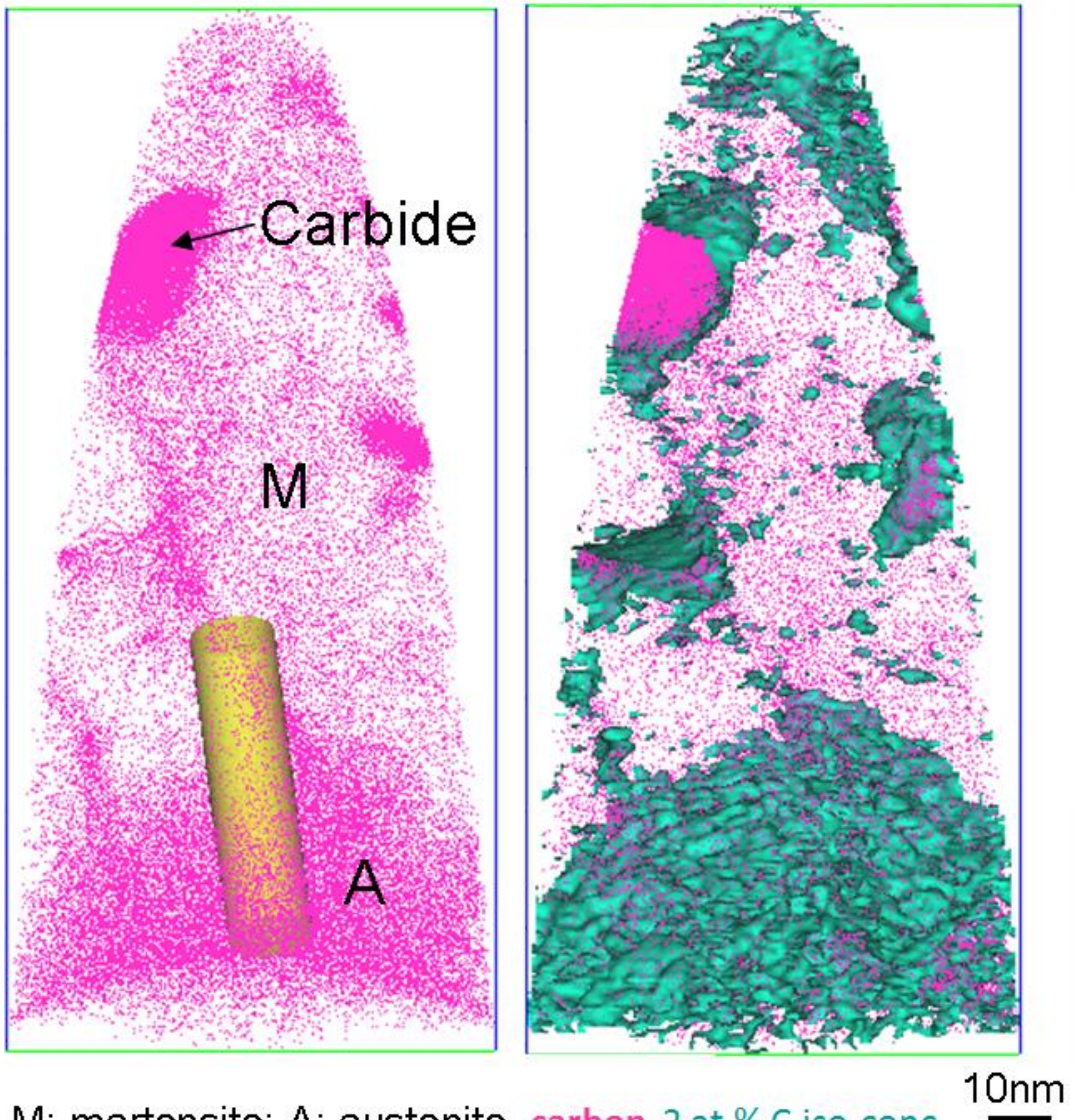

M: martensite; A: austenite carbon 2 at.% C iso-conc.

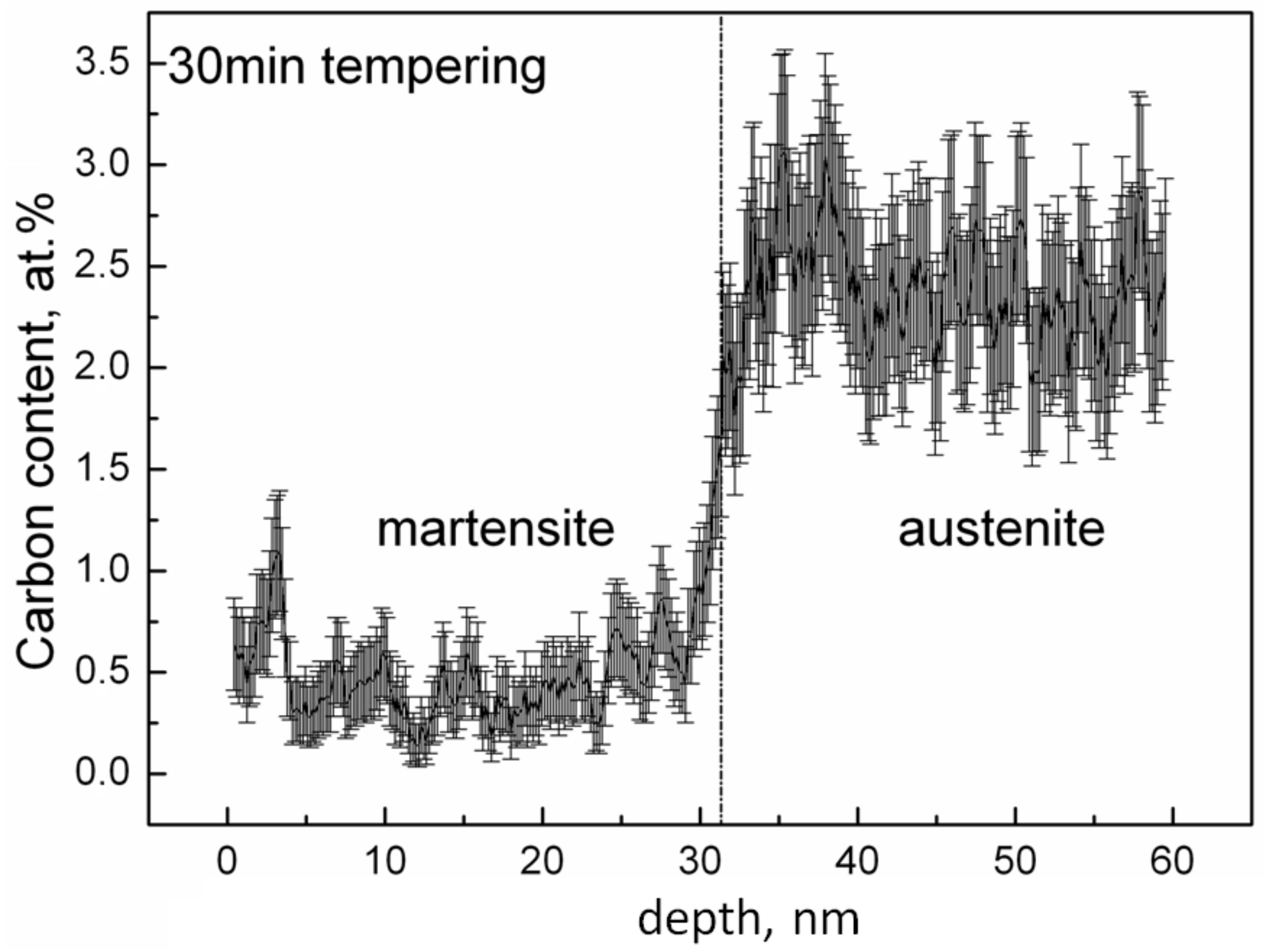

(c)

Fig. 8 (a) 3-D reconstructions (frame scale in nm) of sample after water quenching; The data clearly show that carbon redistribution already occurs during quenching. Cr redistribution does not occur. (b) tempered at 400°C for 1 minute; (c) tempered at 400°C for 30 minutes. Carbon atoms are displayed pink. The different phases are marked in the pictures. Carbon iso-concentration surface (2 at.%, corresponding to 0.44 wt.%, green) and concentration profiles across the phase boundaries along the yellow cylinder) are also shown.

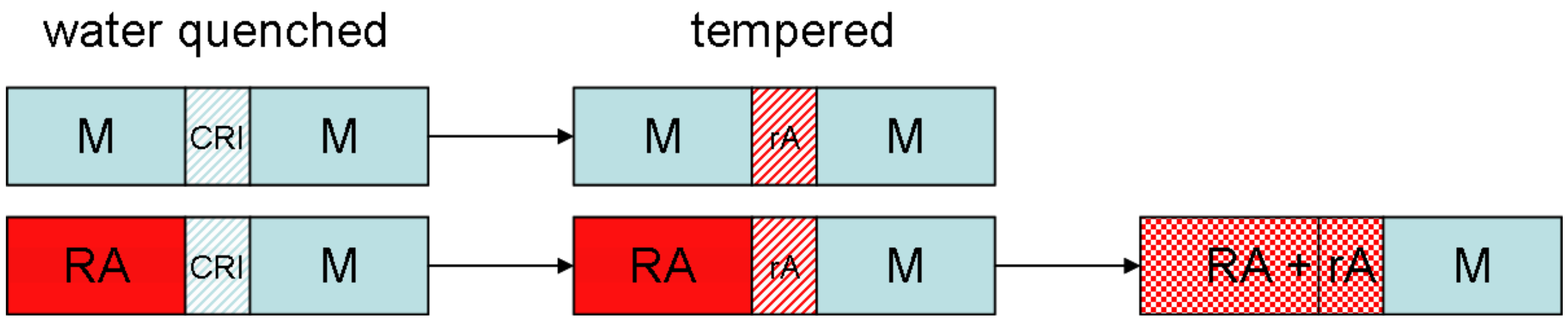

Fig. 9a Schematic illustration of austenite reversion. M: martensite; CRI: carbon-rich interface; RA: retained austenite as obtained after quenching with equilibrium austenite carbon content; rA: reverted austenite formed during 300°C-500°C tempering at interfaces owing to the higher local carbon conent; After sufficient long diffusion time the carbon content in both types of austenite becomes similar.

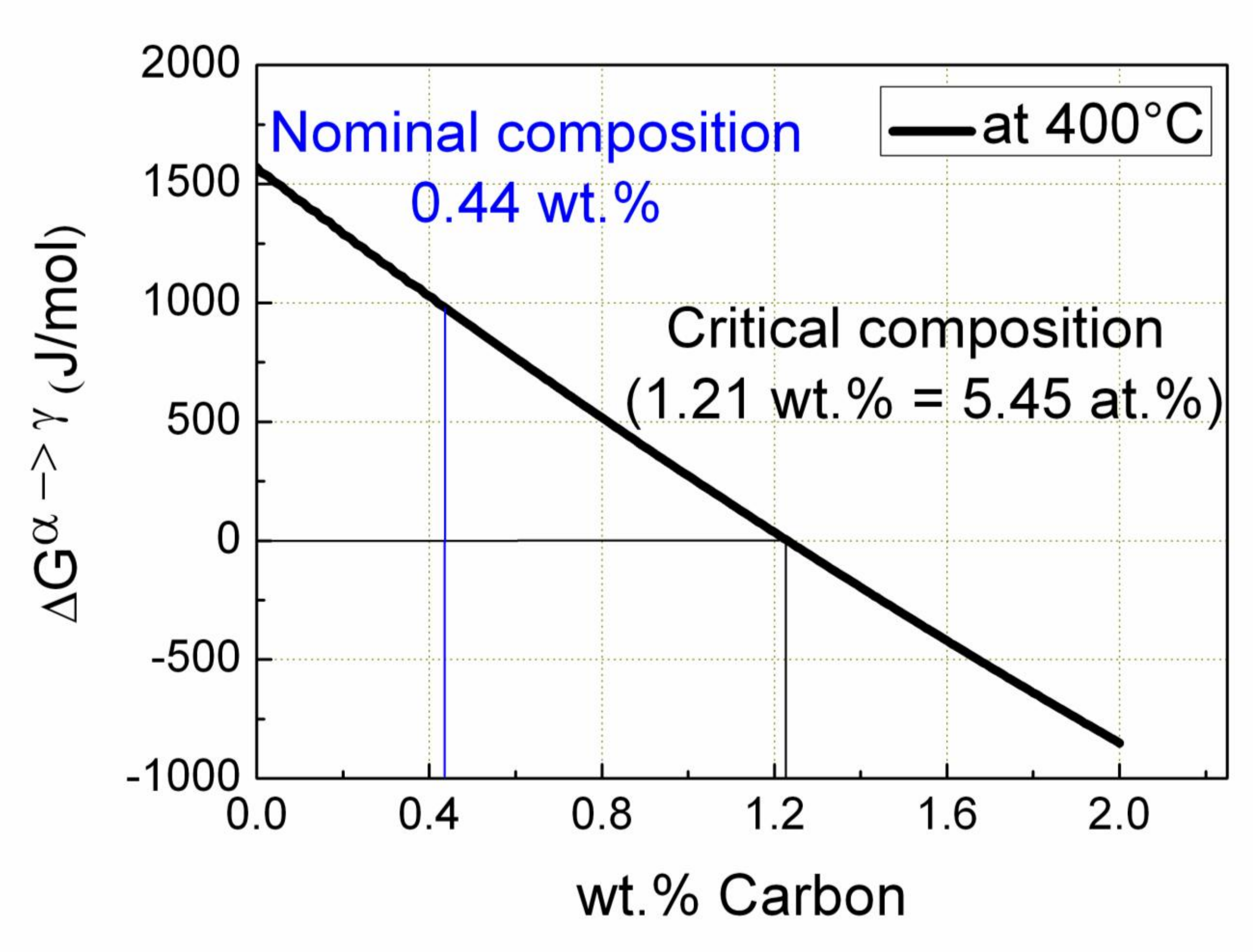

Fig. 9b Calculated driving force for austenite reversion at 400°C (Thermo-Calc TCFE5).

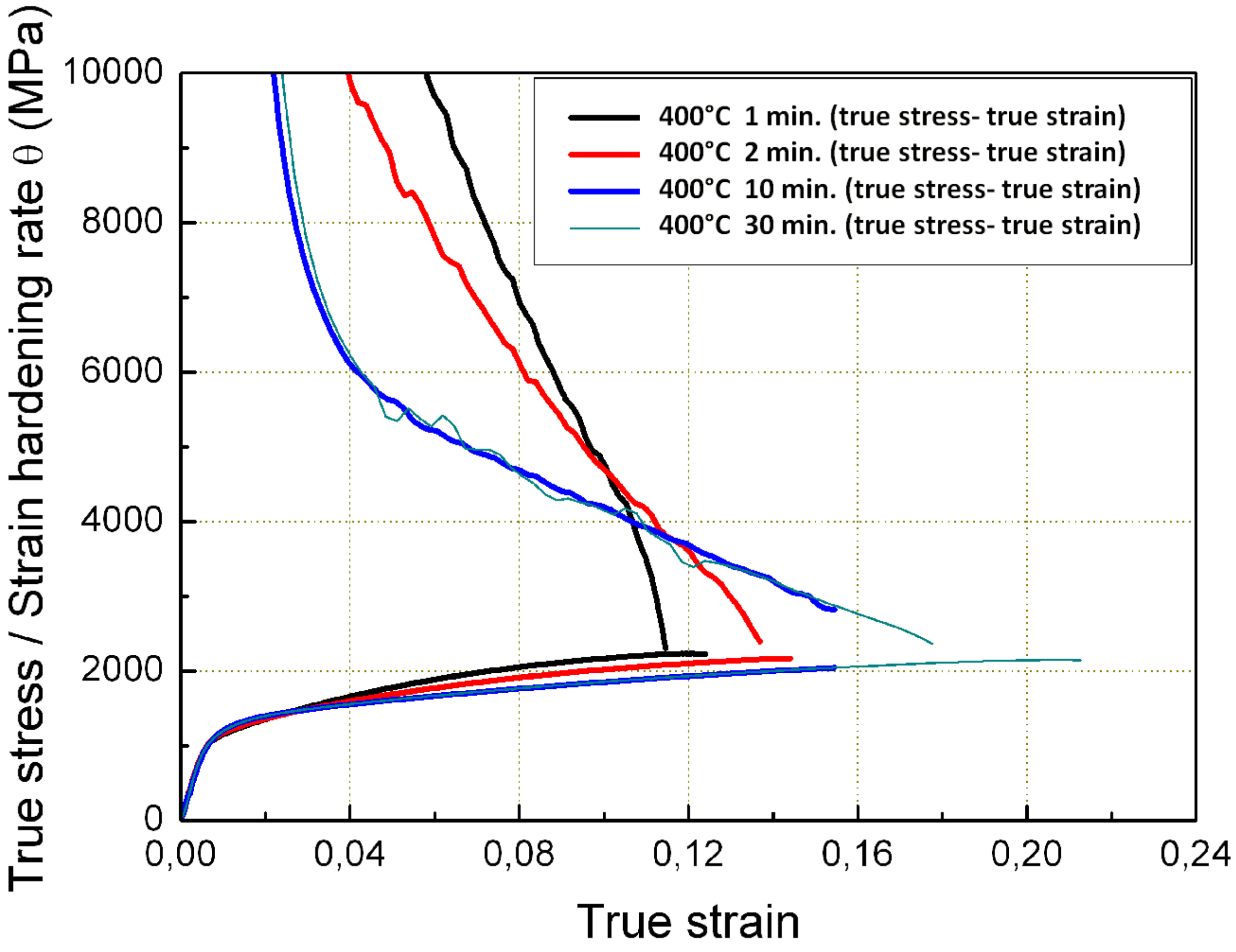

Fig. 10 True stress-true strain curves and corresponding strain hardening curves for the steel after 400°C heat treatment at different times. The data reveal that the tempering, associated with the increase in the austenite content via austenite reversion, yields higher strain hardening reserves at the later stages of deformation due to the TRIP effect.

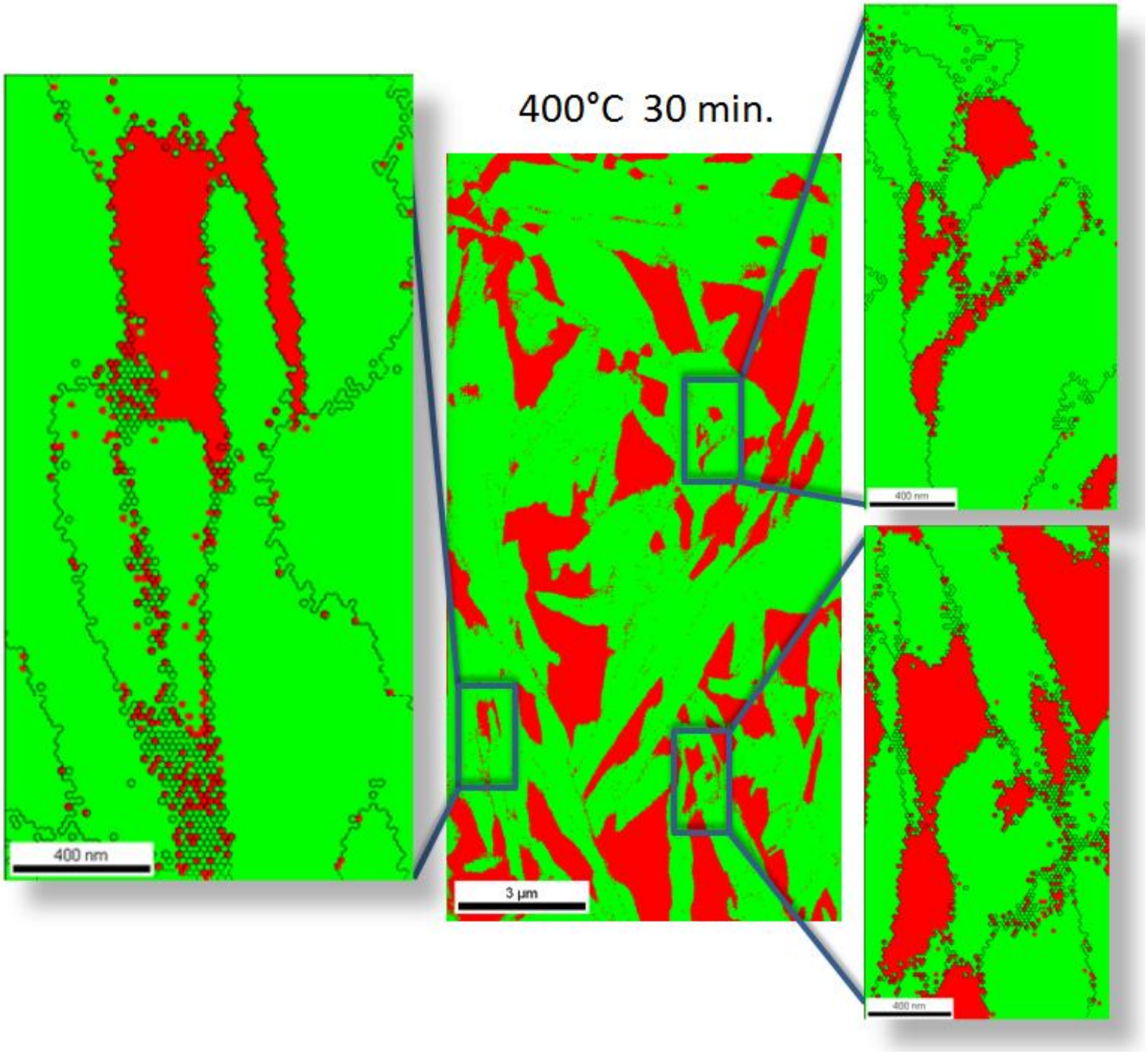

Fig. 11 High resolution EBSD map (20 nm step size) of the sample tempered at 400°C for 30 minutes. The map shows that on some martensite grain boundaries a very thin reverted austenite layer exists. This thin austenite seam can act as a compliance or respectively repair layer against damage percolation entering from the martensite grain. Austenite: red; martensite: green.